\documentclass[english,twoside,referee,useAMS,usenatlib]{mn2e}
\usepackage[T1]{fontenc}
\usepackage[latin9]{inputenc}
\usepackage{graphicx}

\makeatletter

\providecommand{\tabularnewline}{\\}

\makeatletter

\title[Integrated parameters of star clusters]{Integrated parameters of star clusters: A comparison of theory and observations }
\author[A. K. Pandey, T. S. Sandhu, R. Sagar and P. Battinelli]{A. K. Pandey$^{1}$\thanks{E-mail:
pandey@aries.res.in (AKP)}, T. S. Sandhu$^{2}$, R. Sagar$^{1}$, P. Battinelli$^{3}$ \\
$^{1}$ Aryabhatta Research Institute of Observational Sciences, Manora Peak, Nainital, 263 129, Uttarakhand, India \\
$^{2}$Department of Physics, Punjabi University, Patiala, 147002, India \\
$^{3}$INAF, Osservatorio Astronomico di Roma Viale del Parco Mellini 84, 00136 Roma, Italy}

\makeatother

\usepackage{babel}
\makeatother

\begin{document}

\date{Accepted xxxx}

\maketitle
\pagerange{\pageref{firstpage}--\pageref{lastpage}} \pubyear{2009}

\label{firstpage}

\begin{abstract}
This paper presents integrated magnitude and colours for synthetic
clusters. The integrated parameters have been obtained for the whole
cluster population as well as for the main-sequence (MS) population
of star clusters. We have also estimated observed integrated magnitudes
and colours of MS population of galactic open clusters, LMC and SMC
star clusters. It is found that the colour evolution of MS population
of star clusters is not affected by the stochastic fluctuations, however
these fluctuations significantly affect the colour evolution of the
whole cluster population. The fluctuations are maximum in $(V-I)$
colour in the age range 6.7 $<$ log (age) $<$ 7.5. Evolution of
integrated colours of MS population of the clusters in the Milky Way,
LMC and SMC, obtained in the present study are well explained by the
present synthetic cluster model. The observed integrated $(B-V)$
colours of MS population of LMC star clusters having age $\geq$ 500
Myr seem to be distributed around $Z=$ 0.004 model, whereas $(V-I)$
colours are found to be more bluer than those predicted by the $Z=$
0.004 model. $(V-I)$ vs $(B-V)$ two-colour diagram for the MS population
of the Milky Way star clusters shows a fair agreement between the
observations and present model, however the diagrams for LMC and SMC
clusters indicate that observed $(V-I)$ colours are relatively bluer.
Possible reasons for this anomaly have been discussed. Comparison
of synthetic $(U-B)$ vs $(B-V)$ relation with the observed integrated
parameters of whole cluster population of the Milky Way, LMC and SMC
star clusters indicates that majority of the bluest clusters ($(B-V)_{o}<$
0.0) follow the MS population relation. The colour evolution of young
Milky Way, LMC and SMC clusters clusters (6.5 $\leq$ log (age) $\leq$
8.0) also indicates that a large number of young clusters follow the
MS population relation. Therefore, in the absence of a careful
modeling of stochastic effects, age determination of young star clusters
by comparing their integrated colours with whole cluster synthetic
colours may yield erroneous results.
\end{abstract}
\begin{keywords} Open clusters and associations: galaxies-- star
clusters: integrated parameters \end{keywords}

\section{Introduction}

Star clusters are useful objects to test the theories of stellar evolution
and stellar dynamics. The star clusters in the Milky Way and in the
Magellanic Clouds span a wide range in age (from few Myr to few Gyr).
The large range in the age of the clusters allows to observe star
clusters at various epochs in their evolution and make it possible
to identify evolutionary trends.

In spite of the advent of new generation ground and space based telescopes,
the integrated parameters of star clusters are the only observable
parameters to investigate the evolutionary history of stellar systems
beyond the local group of galaxies. In order to interpret the integrated
parameters of extra-galactic star clusters, it is necessary to study
the integrated parameters of star clusters of our galaxy where observation
of individual stars in the cluster region can be carried out to study
the various parameters like age, mass, metallicity, etc. with sufficient
accuracy (e.g. Hancock et al. 2008 and references therein).

Various efforts both from a theoretical or an observational point
of view (see e.g. Searle et al. 1980, Sagar et al. 1983, Chiosi et
al. 1988, Pandey et al. 1989, Battinelli et al. 1994, Brocato et al.
1999, Lata et al. 2002, Bruzual \& Charlot 2003) have been
made to interpret the integrated photometric colours of a simple stellar
population (SSP), like star clusters, in terms of stellar ages and
chemical composition. Lata et al. (2002) have calculated integrated
magnitudes and colours of 140 open clusters of the Milky Way which
in combination with earlier estimates provide integrated photometric
parameters for 352 star clusters. Lata et al. (2002) have reported
for the first time integrated $(V-I)$ and $(V-R)$ colours for 58
and 23 star clusters respectively. Although the integrated $U,B,V$
magnitudes reported by them are for the whole population (main sequence
(MS)+ evolved cluster population), the integrated $R$ and $I$ magnitudes
for clusters younger than 100 Myr are mainly for MS population as
CCD observations of bright stars are not available in these clusters.
Therefore the observed evolution of $(V-R)$ and $(V-I)$ colours
is different from the theoretical evolution (which is for the whole
cluster population) given by Maraston (1998) and Brocato et al. (1999).

Integrated $UBV$ photometry for 147 LMC star clusters was reported
by van den Bergh (1981). The sample of integrated $UBV$ magnitudes
was further enhanced to 624 by Bica et al. (1992, 1996). Recently
Rafelski and Zaritsky (2005,hereafter RZ05) have reported integrated
magnitude and colours for 195 SMC star clusters. A comparison of their
data with the model of Leitherer et al. (1999) and Anders \& Fritz-v.
Alvensleben (2003) indicates a large scatter in the observed data
with a systematic difference between the observed data and the model.

Because of the relevance of the integrated parameters, population
synthesis models have been continuously upgraded over the years (cf.
Brocato et al. 1999, Anders \& Fritz-v. Alvensleben 2003 and references
therein). Similarly the catalogue of the observed integrated parameters
of galactic open clusters, Magellanic Clouds (MC) clusters are frequently
being upgraded. In the present study we have made an attempt to append
the database of integrated parameters of MS population of open clusters,
LMC and SMC clusters. Here we present integrated magnitude and colours
of MS population 66 galactic open clusters, 745 LMC clusters and 238
SMC clusters. The $(B-V)$, $(V-R)$ and $(V-I)$ colours for the
MS population of star clusters in the LMC and SMC are being reported
for the first time. Presently available integrated $(V-R)$ and $(V-I)$
colours of MC star clusters are for the whole population of the cluster.
The integrated colours are frequently used to date the clusters (e.g.
Elson \& Fall 1985, Chiosi et al. 1988, RZ05, Hancock et al. 2008).
Since the integrated colours of the whole cluster population are severely
affected by the stochastic fluctuation (cf. Chiosi et al. 1988; Sec.
2 of present study), the age calibration of the clusters will also
be affected accordingly. Whereas the colour evolution of MS population
is quite systematic, therefore the integrated colours of MS population
should give a better estimate of the age of the clusters. In the present
study we have also calculated integrated parameters for MS and whole
population of the synthetic clusters. The comparison with the observational
data of galactic open clusters and MC star clusters has also been
carried out. The paper is organized as follows. In section Sec. 2
a detailed description of the model is presented. Section 3 describes
the estimation of observed integrated parameters of star clusters.
In sections 4 and 5, a comparison of observed and theoretical integrated
parameters has been carried out. Section 6 describes the conclusion
of the present study.

\section{Theoretical integrated parameters of star clusters}

Various evolutionary population synthesis (EPS) models, as mentioned
in Sec. 1, have been developed for SSPs. The results of EPS may differ
from one another due to input parameters. The comparison of EPS with
observations can give information about acceptability of a particular
EPS (see e.g. Lata et al. 2002).

We have generated synthetic colour-magnitude diagram (CMDs) of open
clusters (Sandhu et al. 2003) using stellar evolutionary models by
Girardi et al. (2002). The synthetic CMDs are constructed
using the technique described by Chiosi et al. (1989). Briefly, this
technique consists of random generation of stars by means of a Monte
Carlo technique and distributing the stars along a given isochrone
according to evolutionary phases and the initial mass function (IMF).
The following expression is used to describe the IMF.

\[
dN=AM^{-X}dM\]
 where $dN$ is the number of stars in the mass interval $dM$, $X$
is the slope of the mass function. The Salpeter (1955) value for the
slope of the mass function is 2.35. The constant $A$ is fixed in
such a way that the initial mass of cluster stars having
masses between 0.6 $\leq$ $M{}_{\odot}$$\leq$ 40 is about 4000
$M{}_{\odot}$ (for details see Sandhu et al. 2003). 
The initial mass value of 4000 $M{}_{\odot}$ is selected as it 
represents approximate average mass of LMC clusters (see e.g. Girardi \& Bica
1993, their figure 8). The contribution
of binary content has not been taken into account. The star formation
is assumed to be instantaneous. The integrated magnitudes and colours
for the whole cluster population (i.e. main sequence and red giants
population) as well as for main sequence population were calculated
using the procedure described by Pandey et al. (1989). One hundred
simulations at log (age) interval of 0.1 were carried out (for various
combinations of the metallicity $Z$ = 0.001, 0.004, 0.008, and 0.02
and mass function slope $X$ = 1.0, 1.35, 2.35, and 3.35) and then
averaged. In the case of $Z$ = 0.001 we have also used the model
by Bertelli et al. (1992) as model by Girardi et al. is available
for log age $\ge$ 7.6. The integrated magnitudes and colours of synthetic
clusters having Salpeter mass function and various assumed values
of metallicity are given in Tables 1 and 2. A sample of the Tables
is shown here. Complete tables are available in electronic form only.

\begin{table}
\caption{Integrated magnitude and colours of synthetic cluster (MS population)
obtained in the present work. The complete table is available in electronic
form only. }

\begin{tabular}{cccccccccccc}
\hline 
{\footnotesize MF Slope}  & Age  & $(U-V)_{0}$  & \emph{err}  & $(U-B)_{0}$  & \emph{err}  & $(B-V)_{0}$  & \emph{err}  & $(V-R)_{0}$  & \emph{err}  & $(V-I)_{0}$  & \emph{err}\tabularnewline
{\footnotesize \& Metallicity}  & \emph{log t}  & mag  & mag  & mag  & mag  & mag  & mag  & mag  & mag  & mag  & mag\tabularnewline
\hline 
{\footnotesize $X$ = 2.35}  & 7.8  & -0.776  & 0.007  & 0.647  & 0.005  & -0.129  & 0.002  & -0.049  & 0.001  & -0.113  & 0.002\tabularnewline
{\footnotesize Z = 0.001}  & 7.9  & -0.714  & 0.005  & -0.599  & 0.004  & -0.115  & 0.002  & -0.042  & 0.001  & -0.097  & 0.002\tabularnewline
 & 8.0  & -0.645  & 0.005  & -0.546  & 0.003  & -0.099  & 0.002  & -0.034  & 0.001  & -0.078  & 0.002\tabularnewline
 & 8.1  & -0.567  & 0.005  & -0.488  & 0.003  & -0.080  & 0.002  & -0.024  & 0.001  & -0.055  & 0.002\tabularnewline
 & 8.2  & -0.520  & 0.005  & -0.449  & 0.003  & -0.071  & 0.002  & -0.020  & 0.001  & -0.044  & 0.002\tabularnewline
 & {\footnotesize .....}  & {\footnotesize .....}  & {\footnotesize .....}  & {\footnotesize .....}  & {\footnotesize .....}  & {\footnotesize .....}  & {\footnotesize .....}  & {\footnotesize .....}  & {\footnotesize .....}  & {\footnotesize .....}  & {\footnotesize .....}\tabularnewline
 & {\footnotesize .....}  & {\footnotesize .....}  & {\footnotesize .....}  & {\footnotesize .....}  & {\footnotesize .....}  & {\footnotesize .....}  & {\footnotesize .....}  & {\footnotesize .....}  & {\footnotesize .....}  & {\footnotesize .....}  & {\footnotesize .....}\tabularnewline
% & {\small .....} &  &  &  &  &  &  &  &  &  & \tabularnewline
{\footnotesize $X$ = 2.35}  & 6.6  & -1.283  & 0.020  & -1.040  & 0.015  & -0.243  & 0.005  & -0.102  & 0.002  & -0.240  & 0.006\tabularnewline
{\footnotesize Z = 0.004}  & 6.7  & -1.259  & 0.020  & -1.023  & 0.015  & -0.237  & 0.005  & -0.100  & 0.002  & -0.234  & 0.005\tabularnewline
 & 6.8  & -1.248  & 0.015  & -1.013  & 0.011  & -0.235  & 0.004  & -0.099  & 0.002  & -0.233  & 0.004\tabularnewline
 & 6.9  & -1.144  & 0.018  & -0.934  & 0.013  & -0.210  & 0.004  & -0.088  & 0.002  & -0.205  & 0.005\tabularnewline
 & 7.0  & -1.126  & 0.015  & -0.920  & 0.011  & -0.206  & 0.004  & -0.086  & 0.002  & -0.201  & 0.004\tabularnewline
 & {\footnotesize .....}  & {\footnotesize .....}  & {\footnotesize .....}  & {\footnotesize .....}  & {\footnotesize .....}  & {\footnotesize .....}  & {\footnotesize .....}  & {\footnotesize .....}  & {\footnotesize .....}  & {\footnotesize .....}  & {\footnotesize .....}\tabularnewline
 & {\footnotesize .....}  & {\footnotesize .....}  & {\footnotesize .....}  & {\footnotesize .....}  & {\footnotesize .....}  & {\footnotesize .....}  & {\footnotesize .....}  & {\footnotesize .....}  & {\footnotesize .....}  & {\footnotesize .....}  & {\footnotesize .....}\tabularnewline
{\footnotesize $X$ = 2.35}  & 6.6  & -1.260  & 0.019  & -1.025  & 0.014  & -0.235  & 0.005  & -0.099  & 0.002  & -0.234  & 0.005\tabularnewline
{\footnotesize {} Z = 0.008}  & 6.7  & -1.243  & 0.019  & -1.011  & 0.014  & -0.232  & 0.005  & -0.098  & 0.002  & -0.231  & 0.005\tabularnewline
 & 6.8  & -1.227  & 0.016  & -0.998  & 0.012  & -0.229  & 0.004  & -0.097  & 0.002  & -0.228  & 0.005\tabularnewline
 & 6.9  & -1.112  & 0.017  & -0.911  & 0.013  & -0.201  & 0.004  & -0.083  & 0.002  & -0.196  & 0.005\tabularnewline
 & 7.0  & -1.097  & 0.010  & -0.899  & 0.008  & -0.198  & 0.002  & -0.082  & 0.001  & -0.192  & 0.003\tabularnewline
 & {\footnotesize .....}  & {\footnotesize .....}  & {\footnotesize .....}  & {\footnotesize .....}  & {\footnotesize .....}  & {\footnotesize .....}  & {\footnotesize .....}  & {\footnotesize .....}  & {\footnotesize .....}  & {\footnotesize .....}  & {\footnotesize .....} \tabularnewline
 & {\footnotesize .....}  & {\footnotesize .....}  & {\footnotesize .....}  & {\footnotesize .....}  & {\footnotesize .....}  & {\footnotesize .....}  & {\footnotesize .....}  & {\footnotesize .....}  & {\footnotesize .....}  & {\footnotesize .....}  & {\footnotesize .....} \tabularnewline
{\footnotesize $X$ = 2.35}  & 6.6  & -1.226  & 0.020  & -1.001  & 0.015  & -0.225  & 0.005  & -0.095  & 0.002  & -0.226  & 0.006\tabularnewline
{\footnotesize Z = 0.02}  & 6.7  & -1.230  & 0.019  & -1.004  & 0.015  & -0.227  & 0.005  & -0.096  & 0.002  & -0.227  & 0.006\tabularnewline
 & 6.8  & -1.201  & 0.011  & -0.982  & 0.009  & -0.219  & 0.003  & -0.092  & 0.001  & -0.219  & 0.003\tabularnewline
 & 6.9  & -1.077  & 0.009  & -0.886  & 0.007  & -0.191  & 0.002  & -0.078  & 0.001  & -0.184  & 0.003\tabularnewline
 & 7.0  & -1.048  & 0.015  & -0.863  & 0.012  & -0.184  & 0.004  & -0.074  & 0.002  & -0.176  & 0.004\tabularnewline
 & {\footnotesize .....}  & {\footnotesize .....}  & {\footnotesize .....}  & {\footnotesize .....}  & {\footnotesize .....}  & {\footnotesize .....}  & {\footnotesize .....}  & {\footnotesize .....}  & {\footnotesize .....}  & {\footnotesize .....}  & {\footnotesize .....} \tabularnewline
 & {\footnotesize .....}  & {\footnotesize .....}  & {\footnotesize .....}  & {\footnotesize .....}  & {\footnotesize .....}  & {\footnotesize .....}  & {\footnotesize .....}  & {\footnotesize .....}  & {\footnotesize .....}  & {\footnotesize .....}  & {\footnotesize .....}\tabularnewline
\hline
\end{tabular}
\end{table}

\begin{table}
\caption{Integrated magnitude and colours of synthetic clusters (whole population)
obtained in the present work. The complete table is available in electronic
form only.}

\begin{tabular}{cccccccccccc}
\hline 
{\footnotesize MF Slope}  & Age  & $(U-V)_{0}$  & \emph{err}  & $(U-B)_{0}$  & \emph{err}  & $(B-V)_{0}$  & \emph{err}  & $(V-R)_{0}$  & \emph{err}  & $(V-I)_{0}$  & \emph{err}\tabularnewline
{\footnotesize \& Metallicity}  & \emph{log t}  & mag  & mag  & mag  & mag  & mag  & mag  & mag  & mag  & mag  & mag\tabularnewline
\hline 
{\footnotesize $X$ = 2.35}  & 7.8  & -0.353  & 0.106  & -0.399  & 0.063  & 0.045  & 0.075  & 0.097  & 0.063  & 0.236  & 0.142\tabularnewline
{\footnotesize {} Z = 0.001}  & 7.9  & -0.288  & 0.115  & -0.353  & 0.062  & 0.065  & 0.073  & 0.107  & 0.059  & 0.258  & 0.130\tabularnewline
 & 8.0  & -0.232  & 0.105  & -0.310  & 0.047  & 0.078  & 0.072  & 0.113  & 0.059  & 0.269  & 0.130\tabularnewline
 & 8.1  & 0.031  & 0.249  & -0.214  & 0.064  & 0.244  & 0.191  & 0.251  & 0.163  & 0.559  & 0.340\tabularnewline
 & 8.2  & 0.138  & 0.280  & -0.164  & 0.083  & 0.302  & 0.204  & 0.293  & 0.171  & 0.647  & 0.354\tabularnewline
 & {\footnotesize .....}  & {\footnotesize .....}  & {\footnotesize .....}  & {\footnotesize .....}  & {\footnotesize .....}  & {\footnotesize .....}  & {\footnotesize .....}  & {\footnotesize .....}  & {\footnotesize .....}  & {\footnotesize .....}  & {\footnotesize .....} \tabularnewline
 & {\footnotesize .....}  & {\footnotesize .....}  & {\footnotesize .....}  & {\footnotesize .....}  & {\footnotesize .....}  & {\footnotesize .....}  & {\footnotesize .....}  & {\footnotesize .....}  & {\footnotesize .....}  & {\footnotesize .....}  & {\footnotesize .....}\tabularnewline
{\footnotesize $X$ = 2.35}  & 6.6  & -1.352  & 0.008  & -1.091  & 0.006  & -0.261  & 0.002  & -0.111  & 0.001  & -0.261  & 0.003\tabularnewline
{\footnotesize {} Z = 0.004}  & 6.7  & -1.332  & 0.011  & -1.078  & 0.007  & -0.254  & 0.004  & -0.108  & 0.002  & -0.256  & 0.004\tabularnewline
 & 6.8  & -0.859  & 0.289  & -0.722  & 0.212  & -0.137  & 0.079  & -0.038  & 0.048  & -0.090  & 0.117\tabularnewline
 & 6.9  & -0.873  & 0.353  & -0.769  & 0.214  & -0.104  & 0.167  & -0.009  & 0.124  & -0.026  & 0.279\tabularnewline
 & 7.0  & -0.601  & 0.331  & -0.677  & 0.168  & 0.076  & 0.223  & 0.146  & 0.177  & 0.329  & 0.391\tabularnewline
 & {\footnotesize .....}  & {\footnotesize .....}  & {\footnotesize .....}  & {\footnotesize .....}  & {\footnotesize .....}  & {\footnotesize .....}  & {\footnotesize .....}  & {\footnotesize .....}  & {\footnotesize .....}  & {\footnotesize .....}  & {\footnotesize .....} \tabularnewline
 & {\footnotesize .....}  & {\footnotesize .....}  & {\footnotesize .....}  & {\footnotesize .....}  & {\footnotesize .....}  & {\footnotesize .....}  & {\footnotesize .....}  & {\footnotesize .....}  & {\footnotesize .....}  & {\footnotesize .....}  & {\footnotesize .....} \tabularnewline
{\footnotesize $X$ = 2.35}  & 6.6  & -1.346  & 0.009  & -1.089  & 0.006  & -0.257  & 0.002  & -0.109  & 0.001  & -0.260  & 0.003\tabularnewline
{\footnotesize Z = 0.008}  & 6.7  & -1.117  & 0.256  & -0.949  & 0.117  & -0.168  & 0.153  & -0.052  & 0.107  & -0.133  & 0.235\tabularnewline
 & 6.8  & -0.524  & 0.513  & -0.598  & 0.290  & 0.074  & 0.250  & 0.093  & 0.155  & 0.189  & 0.332\tabularnewline
 & 6.9  & -0.827  & 0.413  & -0.817  & 0.187  & -0.010  & 0.261  & 0.063  & 0.191  & 0.125  & 0.418\tabularnewline
 & 7.0  & -0.797  & 0.268  & -0.765  & 0.139  & -0.032  & 0.167  & 0.077  & 0.158  & 0.185  & 0.372\tabularnewline
 & {\footnotesize .....}  & {\footnotesize .....}  & {\footnotesize .....}  & {\footnotesize .....}  & {\footnotesize .....}  & {\footnotesize .....}  & {\footnotesize .....}  & {\footnotesize .....}  & {\footnotesize .....}  & {\footnotesize .....}  & {\footnotesize .....} \tabularnewline
 & {\footnotesize .....}  & {\footnotesize .....}  & {\footnotesize .....}  & {\footnotesize .....}  & {\footnotesize .....}  & {\footnotesize .....}  & {\footnotesize .....}  & {\footnotesize .....}  & {\footnotesize .....}  & {\footnotesize .....}  & {\footnotesize .....}\tabularnewline
{\footnotesize $X$ = 2.35}  & 6.6  & -1.317  & 0.008  & -1.072  & 0.005  & -0.245  & 0.003  & -0.105  & 0.002  & -0.252  & 0.004\tabularnewline
{\footnotesize Z = 0.02}  & 6.7  & -1.083  & 0.237  & -0.925  & 0.117  & -0.158  & 0.130  & -0.049  & 0.086  & -0.127  & 0.186\tabularnewline
 & 6.8  & -0.779  & 0.331  & -0.750  & 0.180  & -0.029  & 0.195  & 0.049  & 0.153  & 0.093  & 0.334\tabularnewline
 & 6.9  & -0.693  & 0.426  & -0.774  & 0.194  & 0.081  & 0.240  & 0.137  & 0.173  & 0.299  & 0.391\tabularnewline
 & 7.0  & -0.640  & 0.321  & -0.789  & 0.115  & 0.149  & 0.211  & 0.254  & 0.169  & 0.601  & 0.386\tabularnewline
 & {\footnotesize .....}  & {\footnotesize .....}  & {\footnotesize .....}  & {\footnotesize .....}  & {\footnotesize .....}  & {\footnotesize .....}  & {\footnotesize .....}  & {\footnotesize .....}  & {\footnotesize .....}  & {\footnotesize .....}  & {\footnotesize .....} \tabularnewline
 & {\footnotesize .....}  & {\footnotesize .....}  & {\footnotesize .....}  & {\footnotesize .....}  & {\footnotesize .....}  & {\footnotesize .....}  & {\footnotesize .....}  & {\footnotesize .....}  & {\footnotesize .....}  & {\footnotesize .....}  & {\footnotesize .....} \tabularnewline
\hline
\end{tabular}
\end{table}

\subsection{The evolution of integrated magnitudes and colours}

\begin{figure*}
\begin{centering}
\includegraphics[width=15cm]{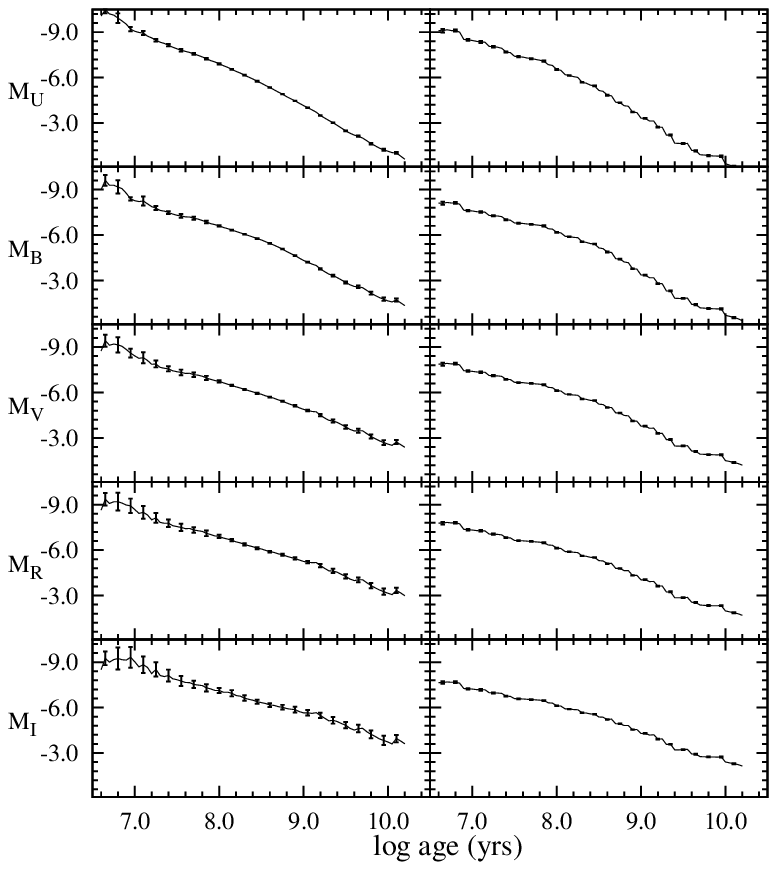} 
\par\end{centering}

\caption{\label{figure:1}Time evolution of integrated $U,B,V,R\,\textup{and}\, I$
magnitudes of a cluster having $X$ = 2.35 and $Z$ = 0.02. Left panel
shows evolution of integrated magnitude of whole cluster population,
whereas right panel shows results for main-sequence stars only.}

\end{figure*}

In Fig. 1 we show the evolution of $U,B,V,R\,\textup{and}\, I$ magnitudes
of a synthetic cluster having solar metallicity $Z=0.02$ and classical
Salpeter mass function $X=2.35$. The error bars show $1\sigma$ dispersion
of the average results obtained from 100 independent simulations.
The $I$ band integrated magnitudes for the synthetic cluster having
evolved stars show largest errors. In both the cases (i.e. whole population
and MS population) integrated luminosity drops in all bands because
of disappearance of bright MS and supergiant stars. The stochastic
fluctuations are relatively less important in the case of MS populations.

\begin{figure*}
\begin{centering}
\includegraphics[width=15cm]{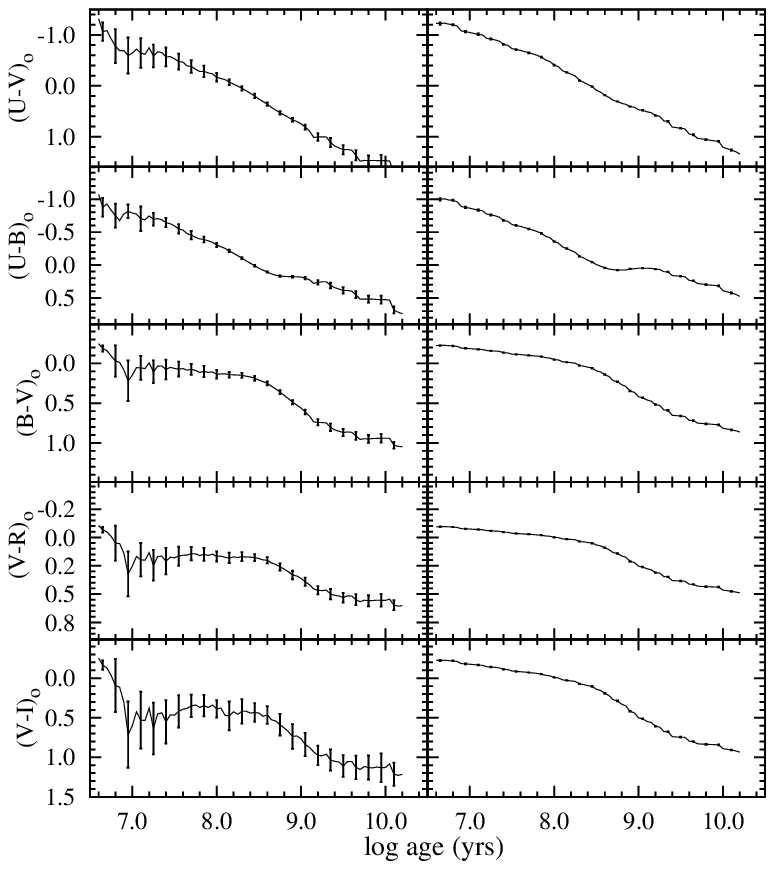} 
\par\end{centering}

\caption{Left \label{fig:2} and right panels show time evolution of integrated
colours of whole cluster population and MS population respectively,
having $X$ = 2.35 and $Z$ = 0.02.}

\end{figure*}

The $(U-B)$, $(U-V)$, $(B-V)$, $(V-R)$ and $(V-I)$ colour evolution
along with standard deviation for the cluster having $X$ = 2.35 and
$Z$ = 0.02 is shown in Fig. 2. All the five colours in the case of
MS population vary smoothly with age, whereas in the case of clusters
having evolved population, only $(U-B)$ and $(U-V)$ colour vary
smoothly with age. The $(B-V)$, $(V-I)$, $(V-R)$ colours show a
plateau around 10 Myr - 200 Myr. In the case of evolved population,
especially in the $(V-I)$ colour, the stochastic fluctuations are
relatively larger than those obtained for MS population only. The
colour variation with the age of the cluster, during 10 Myr to 1000
Myr, is maximum in the $(U-B)$, $(U-V)$ colours, whereas the variation
is minimum for $V-R$ colour.

\subsection{Effect of stochastic fluctuations on colours}

\label{sub:Effect-of-stochastic}

Stochastic effects can produce a significant amount of dispersion
in the integrated colours, especially in the integrated colours of
very young clusters which contain RSG stars (cf. Girardi et al. 1995).
To study the influence of stochastic effects on colour evolution of
synthetic clusters, we carried out one hundred simulations assuming
$Z$ = 0.02 and $X$ = 2.35 and 1.35 for the cluster. We estimated
the mean colour and standard deviation around the mean colours. %
\begin{figure*}
\begin{centering}
\includegraphics[width=12cm]{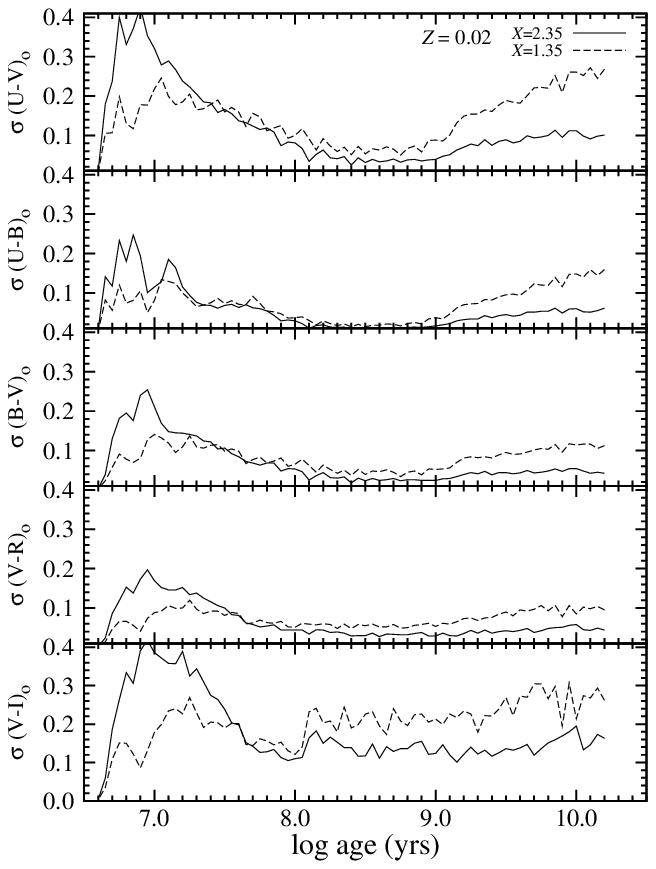} 
\par\end{centering}

\caption{\label{fig:3} Effect of stochastic fluctuations on colours of
the synthetic cluster having post-MS population.}

\end{figure*}

Fig. 3 displays the influence of stochastic fluctuations on integrated
colour as a function of age which indicates that;

i) For log (age) $\leq$ 6.7, the dispersion is low in all the colours.
This is due to lack of evolved stars.

ii) For 6.7 $\leq$ log (age)$\leq$ 7.5, the dispersion is high with
a peak at log (age)$\sim$ 7.0. The large scatter in the colours during
this period is due to small number of RSG stars. The dispersion in
$(V-I)$ is relatively higher at all ages.

iii) In the case of clusters having log (age) $>$ 7.5, the dispersion
in colours decreases with the age.

\subsection{Effect of IMF and chemical composition}

To study the influence of IMF on the integrated magnitude and colours,
we again carried out simulations by varying the slope of the mass
function $X$. %
\begin{figure*}
\begin{centering}
\includegraphics[width=12cm]{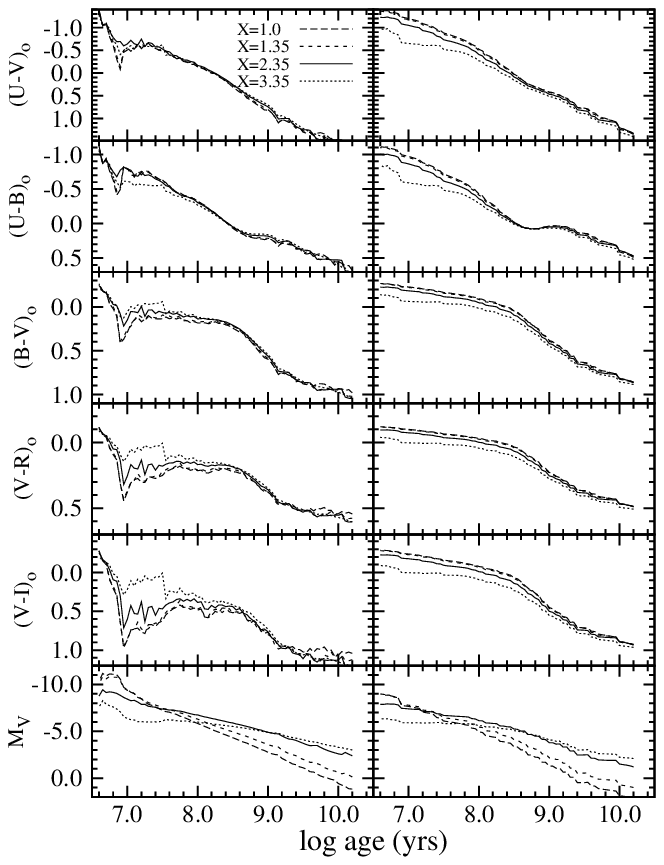} 
\par\end{centering}

\caption{\label{fig:4}Left panel: Evolution of integrated $V$ magnitude and
colours of synthetic clusters (whole population) having $Z$ = 0.02
and $X$ = 1.0, 1.35, 2.35, 3.35. Right panel: Same as left panel
but for MS population.}

\end{figure*}
Fig. 4 shows evolution of synthetic $V$ magnitude and colours for
$X$ = 1.0, 1.35, 2.35, 3.35. For smaller value of $X$ a large fraction
of stars comes from massive stars. Hence, integrated magnitudes of
the cluster become brighter. However for ages greater than 1000 Myr
the larger value of $X$ results in fainter clusters because at ages
greater 1000 Myr the luminosity contribution comes from less massive
(i.e. fainter) stars. The evolution of integrated colours of clusters
having log (age)$\geq$ 7.5 is not affected by the variation of IMF.
However, the colour evolution for log (age) $<$ 7.5 is influenced
by the IMF. The dispersion is maximum in the $(V-I)$ colour of the
whole cluster population. Whereas the effect of IMF is less prominent
on the integrated $(V-I)$ colours of MS population of the clusters.

The influence of metallicity on the integrated magnitude and colours
has been studied assuming $Z$ = 0.004, 0.008, 0.02 and shown in %
\begin{figure*}
\begin{centering}
\includegraphics[width=12cm]{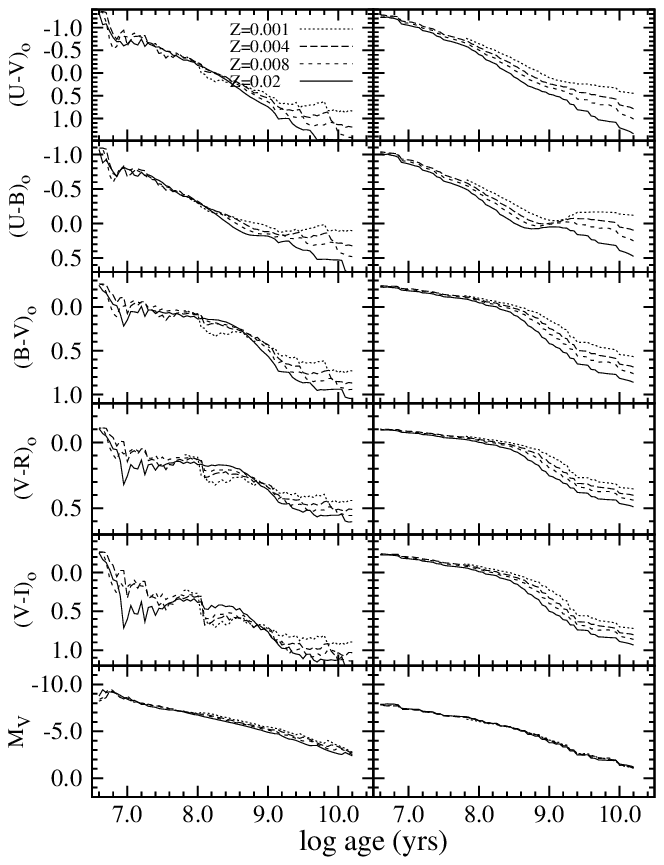} 
\par\end{centering}

\caption{\label{fig:5}Left panel: Effect of metallicity on the integrated
magnitude and colours of synthetic clusters (whole population) having
Salpeter mass function and $Z$ = 0.004, 0.008, 0.02. Right panel:
Same as left panel but for MS population.}

\end{figure*}
Fig. 5. The metallicity variation does not show any significant impact
on the evolution of integrated $V$ magnitude as well as on the $(U-B)$,
$(U-V)$, $(B-V)$, $(V-R)$ and $(V-I)$ colour evolution of clusters
having log (age) $<$8.5, however, for older clusters the colours
become bluer for $Z$ = 0.008 and 0.004 than the solar metallicity
models. The same effect has also been reported and discussed by Brocato
at al. (1999). However, the effect of metallicity on the $(B-V)$
colour obtained by us is not so prominent as reported by Brocato et
al. (1999).

\subsection{Comparison with previous models}

The EPS are being continuously upgraded over the years (e.g. Brocato
et al. 1999, Maraston 1998). The comparison of different EPS models
is not a simple task, because many factors contribute to produce different
results (cf. Maraston 1998). Maraston (1998) and Brocato et al. (1999)
have compared various ESP models and concluded that the evolution
of $(U-B)$ and $(B-V)$ is very similar for all sets of models, whereas
in the case of near $IR$ indices at intermediate ages major source
of discrepancy between the models arises due to Asymptotic Giant Branch
(AGB) phase.

\begin{figure*}
\begin{centering}
\includegraphics[width=12cm]{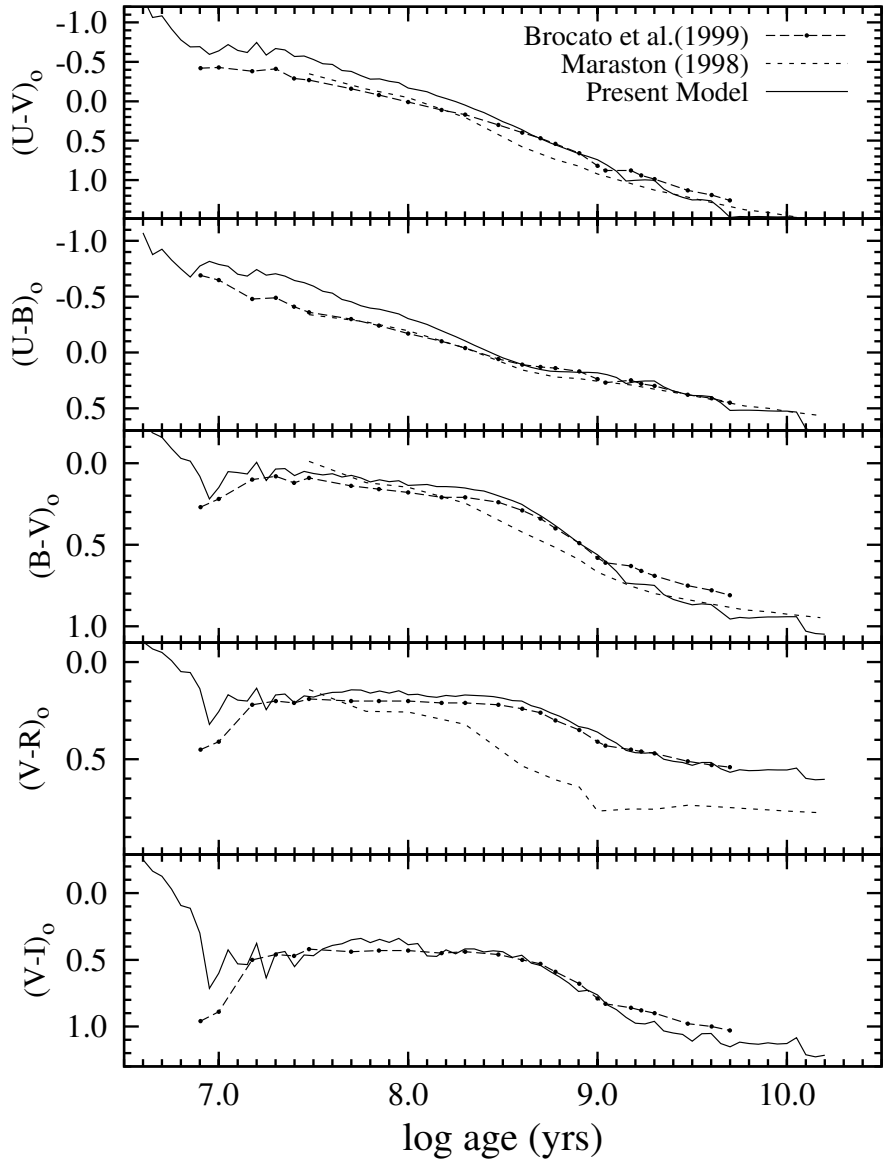} 
\par\end{centering}

\caption{\label{fig:6} Comparison of integrated colours (for $X$ = 2.35,
$Z$ = 0.02) obtained in the present work with those obtained by Brocato
et al. (1999), Maraston (1998).}

\end{figure*}

In Fig. 6 we compare our results for solar metallicity and Salpeter
IMF (whole cluster population) with those by Brocato et al. (1999)
and Maraston (1998). The comparison indicates that;

i) The $(U-B)$ and $(U-V)$ colours obtained in the present work
are slightly bluer in comparison to those given by Brocato et al.
(1999) and Maraston (1998).

ii) An agreement can be seen between $(B-V)$ colours obtained in
the present work and those by Brocato et al. (1999). Keeping in mind
the errors reported in Sec. 2.1 the $(B-V)$ colours obtained in the
present work are in reasonable agreement with those give by Maraston
(1998).

iii) For synthetic clusters having age $>$ 15 Myr an agreement can
be seen between $(V-R)$ colour evolution obtained in present work
and that predicated by Brocato et al. (1999). For clusters younger
than $\sim$ 15 Myr, the $(V-R)$ colours obtained in the present
work are bluer than the colours predicted by Brocato et al. (1999).
On the other hand the evolution of $(V-R)$ colour predicated by Maraston
(1998) does not match either with the present work or with the work
of Brocato et al. (1999). The $(V-R)$ colours by Maraston (1998)
are significantly redder than those obtained in the present work or
by Brocato et al. (1999).

iv) Considering the errors in the present $(V-I)$ colour estimation
(cf. Fig. 2), the predicted $(V-I)$ colour evolution of clusters
in the present work and that by Brocato et al. (1999) is in fair agreement.

\section{Observed integrated parameters}

\label{sec:Observed-integrated-parameters sec 3}

\subsection{Galactic clusters}

Using the observation of individual stars of a galactic star cluster,
integrated photometric parameters have been obtained by several authors
(Lata et al. 2002 and reference therein). Lata et al. (2002) have
obtained integrated parameters of 140 clusters, which in combination
with earlier estimates provide integrated photometric parameters for
352 clusters. Lata et al. (2002) have reported integrated $(V-R)$
and $(V-I)$ colours for the first time, however for most of the younger
clusters (age $\leq$100 Myr), the integrated $(V-R)$ and $(V-I)$
may represent MS population because young clusters have bright stars
for which generally $(V-R)$ and $(V-I)$ CCD observations are not
available. In the present work we have estimated integrated colours
for only $MS$ population of galactic clusters. The $UBV$ Johnson
and $RI$ Cousins CCD data along with the distance, $E(B-V)$ and
age of the galactic clusters have been taken from WEBDA database.
The integrated magnitude and colours were calculated using the procedure
described by Pandey et al. (1989). The colour excesses $E(U-B),E(U-V),E(V-R)$
and $E(V-I)$ have been calculated from $E(B-V)$ using the relations
$E(U-B)=0.72E(B-V);E(U-V)=1.72E(B-V);E(V-R)=0.60E(B-V)$ and $E(V-I)=1.25E(B-V)$.
The possible source of errors in determination of the integrated parameters
are same as described by Sagar et al. (1983). The uncertainty in estimation
of integrated magnitude and colours is $\sim$ 0.5 mag and $\sim$0.2
mag, respectively. Battinelli et al. (1994) have also reported the
same order of uncertainty in their estimation of integrated parameters.
A comparison of integrated parameters obtained by various authors
has been carried out by Lata et al. (2002, their figure 1), which
clearly supports the reported errors of 0.5 mag in estimation of observed
integrated magnitudes.

Table 3 gives a sample of the catalogue of intrinsic integrated $M_{V}$
magnitude and $(U-B)$, $(U-V)$, $(B-V)$, $(V-R)$ and $(V-I)$
colours. The complete catalogue is available in electronic form only.

\begin{table*}
\centering %
\begin{minipage}[c][1\totalheight]{140mm}%
\caption{ Observed MS integrated magnitude and colours of galactic clusters
obtained in the present work. The complete table is available in electronic
form only.}

\begin{tabular}{@{}rrrrrrrrrrr@{}}
\hline 
\medskip{}

Cluster  & $(m-M)$  & E(B-V)  & Age  & $M_{V}$  & $(U-V)_{0}$  & $(U-B)_{0}$  & $(B-V)_{0}$  & $(V-R)_{0}$  & $(V-I)_{0}$  & \tabularnewline
 & (mag)  & (mag)  & log \textit{t}  & (mag)  & (mag)  & (mag)  & (mag)  & (mag)  & (mag)  & \tabularnewline
\hline 
Be 20  & 15.00  & 0.12  & 9.70  & -0.88  & --  & --  & 0.53  & 0.31  & 0.63  & \tabularnewline
Be 42  & 11.30  & 0.76  & 9.30  & 2.82  & 0.65  & 0.04  & 0.62  & 0.33  & --  & \tabularnewline
Be 64  & 16.20  & 1.05  & 9.00  & -4.33  & 0.23  & 0.01  & 0.23  & 0.02  & 0.13  & \tabularnewline
Be 69  & 14.30  & 0.65  & 9.00  & -2.15  & 0.29  & 0.07  & 0.21  & 0.09  & 0.29  & \tabularnewline
Be 86  & 12.60  & 0.70  & 7.00  & -5.42  & -0.96  & -0.78  & -0.18  & 0.06  & --  & \tabularnewline
\hline
\end{tabular}%
\end{minipage}
\end{table*}

\subsection{Clusters in the Magellanic Cloud}

Integrated $UBV$ photometry for 147 LMC clusters has been reported
by van den Bergh (1981). Bica et al. (1992, 1996) extended the sample
of LMC clusters to a total of 624 objects. Their sample includes fainter
clusters and they claim that the catalogue is complete up to V$\approx$13.2.
Recently Hunter et al. (2003) have studied integrated properties of
939 star clusters in the MC, which were based on ground based CCD
images in $UBVR$ passbands. All of the studies mentioned above are
based on integrated photometry of the clusters, therefore the integrated
parameters are for the whole population (i.e. MS+giant stars) of the
clusters. To our knowledge none of the studies is available in literature
where integrated magnitude and colours for only MS population of MC
star clusters are reported. The integrated colours are being used
since a long time to date the extragalactic clusters (e.g. Girardi
2001 and reference therein, Hunter et al. 2003). As we have discussed
in Sec. 2, the colour evolution of entire population of the cluster
(i.e. MS+giants) is significantly influenced by the stochastic fluctuations
in comparison to the colour evolution of MS population of the cluster,
therefore age of clusters derived from the whole population integrated
colours must be subject to a greater uncertainty in comparison to
those obtained by using the only MS population sample.

The Optical Gravitational Lensing Experiment (OGLE) has reported $BVI$
photometry for 745 LMC clusters (Pietrzynski et al. 1999) and 238
SMC clusters (Pietrzynski et al. 1998). We have used above mentioned
catalogues to calculate the integrated parameters of MS population
of the LMC and SMC clusters. The selection of data sample representing
the main-sequence of MC clusters is an arduous task. The width of
the observed MS depends on the presence binaries, photometric errors,
intra-cluster reddening and spread in metallicity. Presence of equal
mass binaries can reddened the distribution by $\sim0.1$ mag (see
Sandhu et al. 2003). Udalski et al (1998) have reported an error of
$\sim0.05$ mag in the estimation of colours. Since $E(B-V)$ for
MC clusters varies from $\sim0.05$ to $\sim0.15$, an average $E(B-V)$
value may introduce an error of $\sim0.1$ mag in dereddened colours.
The parameters discussed above can broaden the observed MS by $\sim\pm0.15$
mag. To select a MS sample we plotted blue and red envelopes, having
a width of 0.3 mag, around the MS as shown in %
\begin{figure*}
\includegraphics[width=10cm]{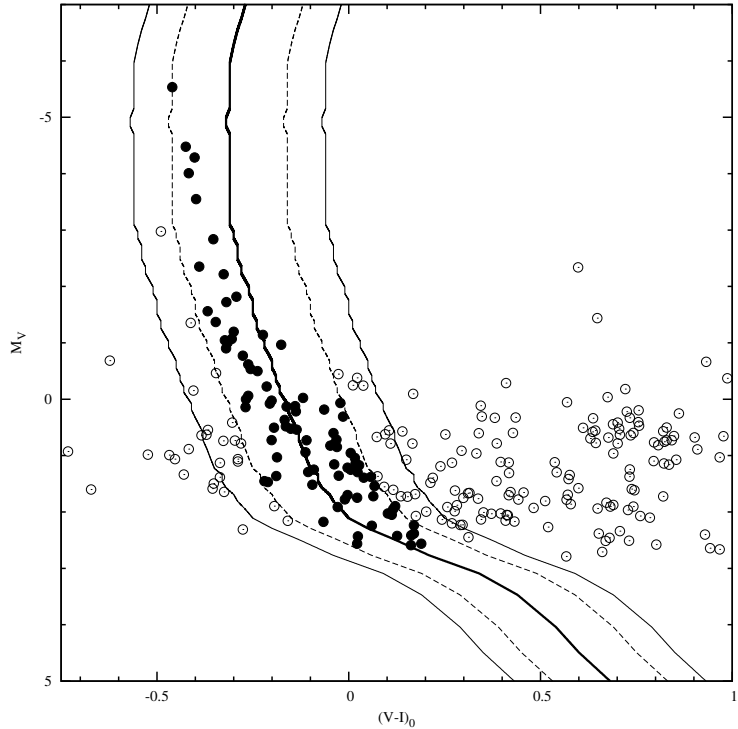}

\caption{\label{fig:7}The blue and red envelopes around the MS (thick curve)
in the case of LMC 327. The width of the envelope in colour is 0.3
mag (dashed curves) and 0.5 mag (thin curves). }

\end{figure*}
Fig. 7. We found that the selected width (0.3 mag) for
MS may exclude about 10 -15\% stars towards brighter end, whereas
blue envelope of the MS excludes about 50\% stars towards fainter
end ($M_{V}~>1$). An increase in width of the MS up to 0.5 mag in
colour includes almost all the brighter stars ($M_{V}~>1$) in the
sample, whereas towards fainter end the blue envelope includes about
80 - 90\% stars blueward of the MS. Further increase in the width
of the MS includes insignificant number of stars towards the bluer
side of the MS, but includes more stars towards the red side of the
MS. Therefore, we select the width of the MS as 0.5 mag. A further
broader MS will have a higher probability to include non-MS stars.
%Non inclusion of $\sim10-20\%$ fainter stars will not affect the estimation of integrated parameters significantly.}

The integrated parameters were calculated using a distance modulus of 18.54 mag and 18.93 mag for the LMC and SMC respectively (Keller et al. 2006). For the age range log (age) $\leq$ 7.3, 7.3 $\leq$ log (age) $\leq$ 8.4, and log (age) $>$
8.4, the reddening $E(B-V)$ is assumed to be 0.14 mag, 0.08 mag and
0.03 mag for LMC clusters. For the age range log (age) $\leq$ 7.3,
7.3 $\leq$ log (age) $\leq$ 8.4, and log (age) $>$ 8.4, the reddening
$E(B-V)$ is assumed to be 0.1 mag, 0.08 mag and 0.03 mag for SMC
clusters. The stars above the turnoff points are not considered for
estimating the integrated parameters. All the probable MS stars of the cluster region lying within the width of 0.5 mag were used to calculate the integrated parameters by summing the flux of each star.The catalogues of integrated
parameters of LMC and SMC clusters are given in Tables 4 and 5, respectively.
A sample of these tables are shown here while the complete catalogues
are available in electronic form . The clusters having MS members
less than 10 stars have not been included in the catalogue. The ages
of the LMC and SMC clusters are taken from Pietrzynski et al. (1999)
and Pietrzynski et al. (1998) respectively.

\begin{table*}
\centering %
\begin{minipage}[c][1\totalheight]{140mm}%
\caption{ Observed MS integrated magnitude and colours of LMC clusters obtained
in the present work. N is the number of stars used to calculate the
integrated parameters. The complete table is available in electronic
form only.}

\begin{tabular}{@{}rrrrrrrrrrr@{}}
\hline 
\medskip{}

OGLE-ID  & Other name  & Age  & $M_{V}$  & $(V-I)_{0}$  & $(B-V)_{0}$  & N  &  &  &  & \tabularnewline
 &  & (log \textit{t)}  & (mag)  & (mag)  & (mag)  &  &  &  &  & \tabularnewline
\hline 
LMC0001  & HS81  & 8.33  & -2.71  & 0.04  & --  & 20  &  &  &  & \tabularnewline
LMC0003  & BSDL403  & 8.70  & -1.17  & 0.11  & --  & 13  &  &  &  & \tabularnewline
LMC0004  & H88-85  & 7.60  & -2.60  & -0.01  & --  & 25  &  &  &  & \tabularnewline
LMC0005  & HS83  & 8.15  & -3.80  & -0.06  & --  & 42  &  &  &  & \tabularnewline
\hline
\end{tabular}%
\end{minipage}
\end{table*}

\begin{table*}
\centering %
\begin{minipage}[c][1\totalheight]{140mm}%
\caption{ Observed MS integrated magnitude and colours of SMC clusters obtained
in the present work. N is the number of stars used to calculate the
integrated parameters. The complete table is available in electronic
form only.}

\begin{tabular}{@{}rrrrrrrrrrr@{}}
\hline 
\medskip{}

OGLE-ID  & Other name  & Age  & $M_{V}$  & $(V-I)_{0}$  & $(B-V)_{0}$  & N  &  &  &  & \tabularnewline
 &  & (log \textit{t)}  & (mag)  & (mag)  & (mag)  &  &  &  &  & \tabularnewline
\hline 
%SMC0001 & H86-35  & --  & -1.14  & -0.07  & -0.02  & 9  &  &  &  & \tabularnewline
SMC0002  & HW11  & 8.40  & -4.25  & -0.01  & -0.06  & 149  &  &  &  & \tabularnewline
SMC0003  & L19  & 9.00  & -2.96  & 0.36  & 0.22  & 133  &  &  &  & \tabularnewline
SMC0004  & B10  & --  & -3.88  & -0.04  & -0.13  & 126  &  &  &  & \tabularnewline
SMC0005  & OGLE  & --  & -1.01  & 0.09  & 0.09  & 12  &  &  &  & \tabularnewline
\hline
\end{tabular}%
\end{minipage}
\end{table*}

In Fig. 8 we have plotted magnitude $M_{V}$ of LMC clusters common
in the catalogue of Bica et al. (1992, 1996) and in the present work
along with the model predictions obtained in the present work. The
apparent $V$ magnitudes by Bica et al. (1992, 1996) are converted
to absolute magnitudes by using a distance modulus of 18.54 and $E(B-V)$
= 0.1 mag. Figure 8 shows that the observations are fairly represented
by the model predictions. However, a few clusters having $M_{V}\sim-9.0$
to $-7.0$ mag (whole population; Bica et al. 1992, 1996) show a large
deviation in the sense that $M_{V}$ estimations for MS populations
are too small ($M_{V}\sim-4.0$ to $-1.0$ mag). It is noticed that
most of these clusters are old and have log age $\ge$ 9.0. In estimation
of MS population integrated magnitudes and colours we have excluded
bright evolved stars above turn-off, where as in the whole population
integrated parameters (by Bica et al.) the possibility of bright field
star contamination is higher. Moreover, the stochastic fluctuations
increase for clusters older than log age $\sim$ 9.0. However, a comparison
of observed $M_{V}$ magnitudes of SMC clusters obtained by RZ05 and
those obtained in the present work with the model predictions shown
in Fig. 9, indicates a rather satisfactory agreement. The apparent
$V$ magnitudes by RZ05 are converted to absolute magnitudes by using
a distance modulus of 18.93 and $E(B-V)$ = 0.1 mag. A comparison
of observed colours of LMC and SMC clusters with the model predictions
is shown in Figs 10 and 11 which indicates a fair agreement between
observations and model predictions. Majority of the observational
data points lie within ~0.15 mag from the model predictions. This
can be considered as the maximum uncertainty in the observed colours,
i.e. a combined uncertainty in observed MS colours and whole population
colours. If we consider that both the samples have same order of uncertainty,
then each sample will have an uncertainty of ~0.1 mag. 

\begin{figure*}
\begin{centering}
\textbf{\includegraphics[width=8cm,angle=270]{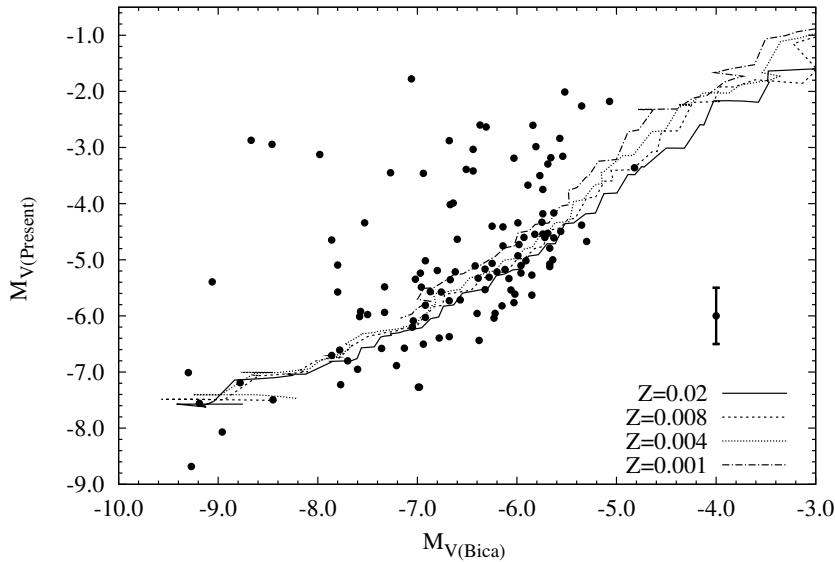}} 
\par\end{centering}

\caption{\textbf{\label{fig:8}}Comparison of observed integrated MS $M_{V}$
magnitudes of LMC clusters obtained in the present work (MS population) and those
given by Bica et al. (1996) for whole cluster population with the present
model predictions. The curves represent the model predictions for
various metallicities.The typical error in estimation of
integrated MS $M_{V}$ is also shown.}

\end{figure*}

\textbf{}%
\begin{figure*}
\textbf{\includegraphics[width=8cm,angle=270]{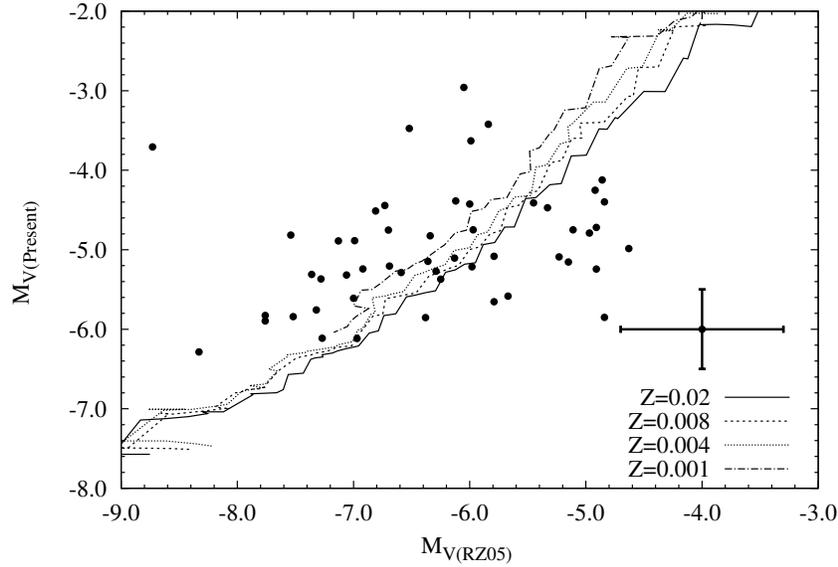} }

\caption{\label{fig:9}Comparison of observed integrated MS $M_{V}$ magnitudes
of SMC clusters obtained in the present work  (MS population) with those given by RZ05
for whole cluster population with the model predictions (continuous
curves). The typical errors in estimation of integrated MS
$M_{V}$ and $M_{V(RZ05)}$ are also shown.}

\end{figure*}

\textbf{}%
\begin{figure*}
\textbf{\includegraphics[width=8cm,angle=270]{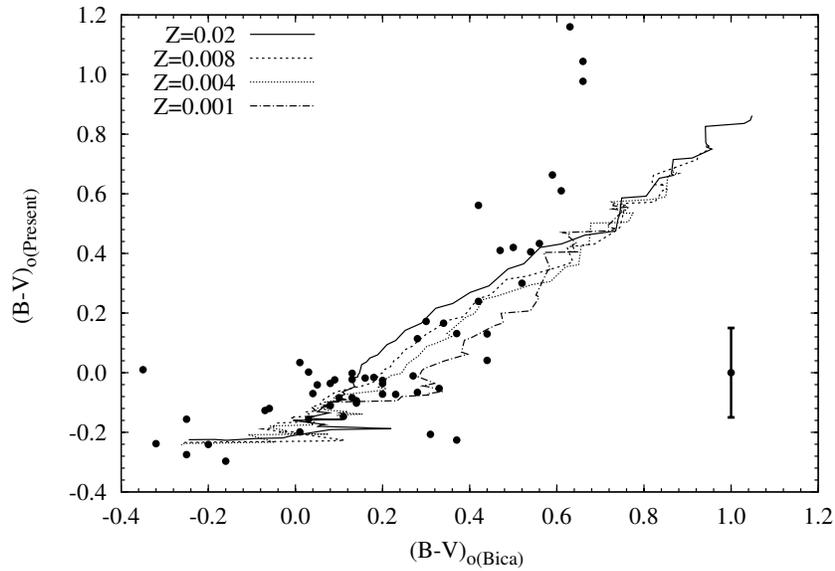} }

\caption{\label{fig:10}Comparison of observed MS integrated $(B-V)$ colours
of LMC clusters obtained in the present work (MS population) and those given by Bica
et al. (1996) for whole cluster population with the model predictions
(continuous curves). The typical error in estimation of integrated
MS $(B-V)$ colours is also shown.}

\end{figure*}

\textbf{}%
\begin{figure*}
\textbf{\includegraphics[width=8cm,angle=270]{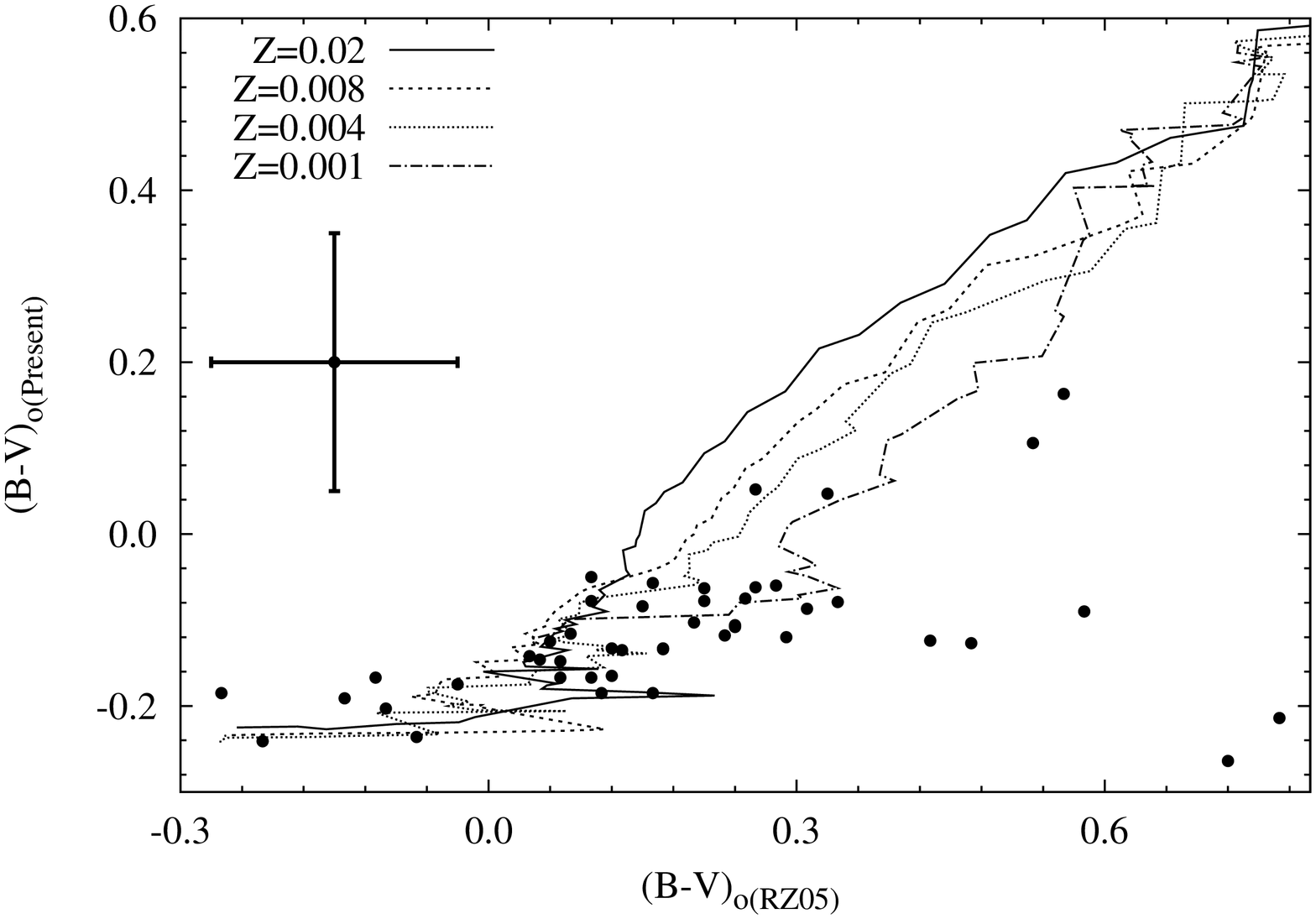} }

\textbf{\includegraphics[width=8cm,angle=270]{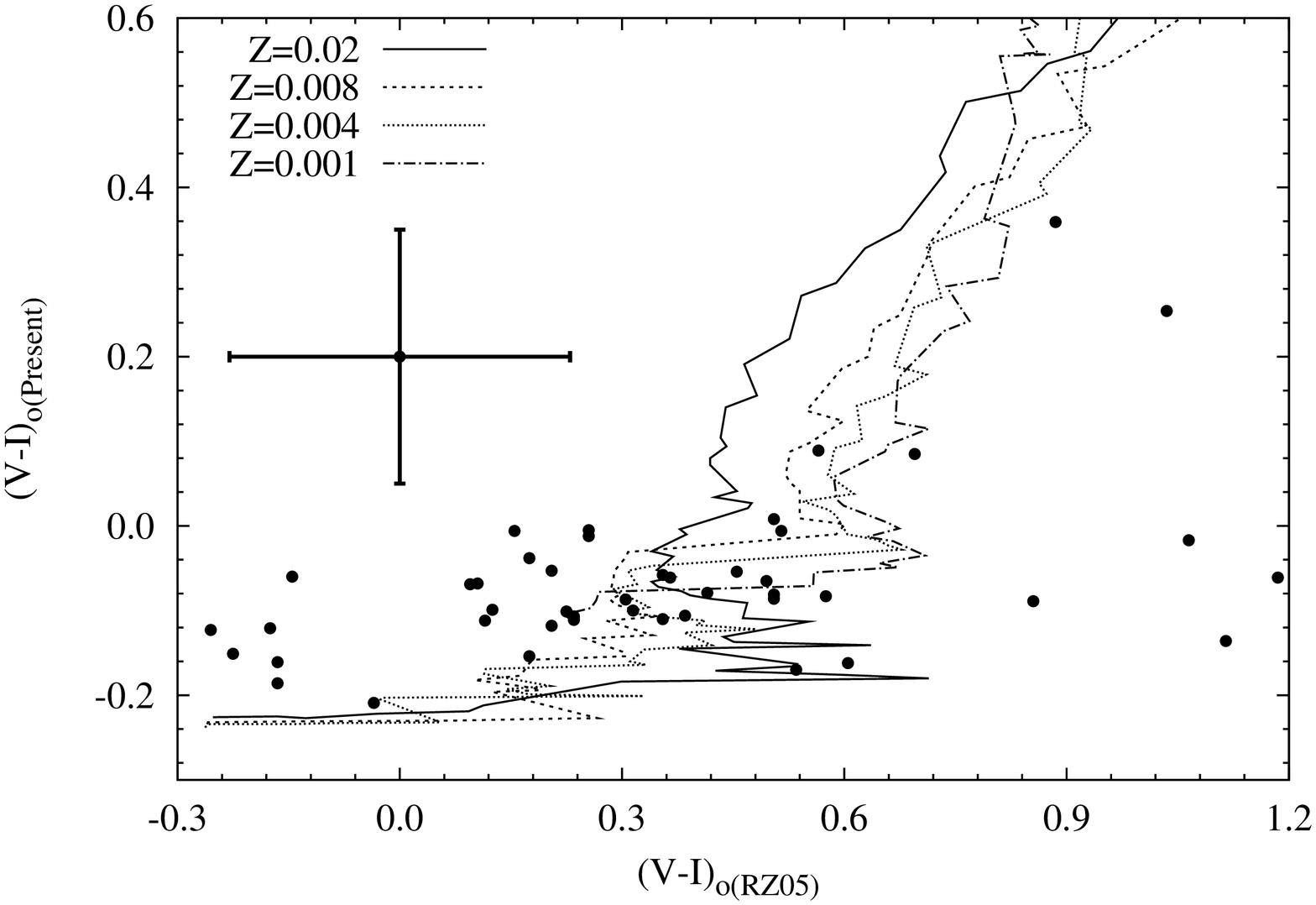} }

\caption{\label{fig:11}Comparison of observed MS integrated colours of SMC
clusters obtained in the present work (MS population) and those given by RZ05 for
whole cluster population with the model predictions (continuous curves).
The typical errors in estimation of integrated MS and whole population
colours are also shown.}

\end{figure*}

\section{Comparison of theoretical predictions with the observations}

\textbf{\label{sec:Comparison-of-theoretical 4}}

The colour of star clusters which form on a short time scale are obvious
choice to test the theoretical models. The star clusters in the Galaxy
and the MC have ages between a few million years to few billion years.
In this section we compare theoretical predictions obtained for synthetic
clusters with the observed integrated parameters of star clusters
in the Milky Way and Magellanic Clouds.

\subsection{Galactic open clusters}

\textbf{}%
\begin{figure*}
\textbf{\includegraphics[angle=270,width=12cm]{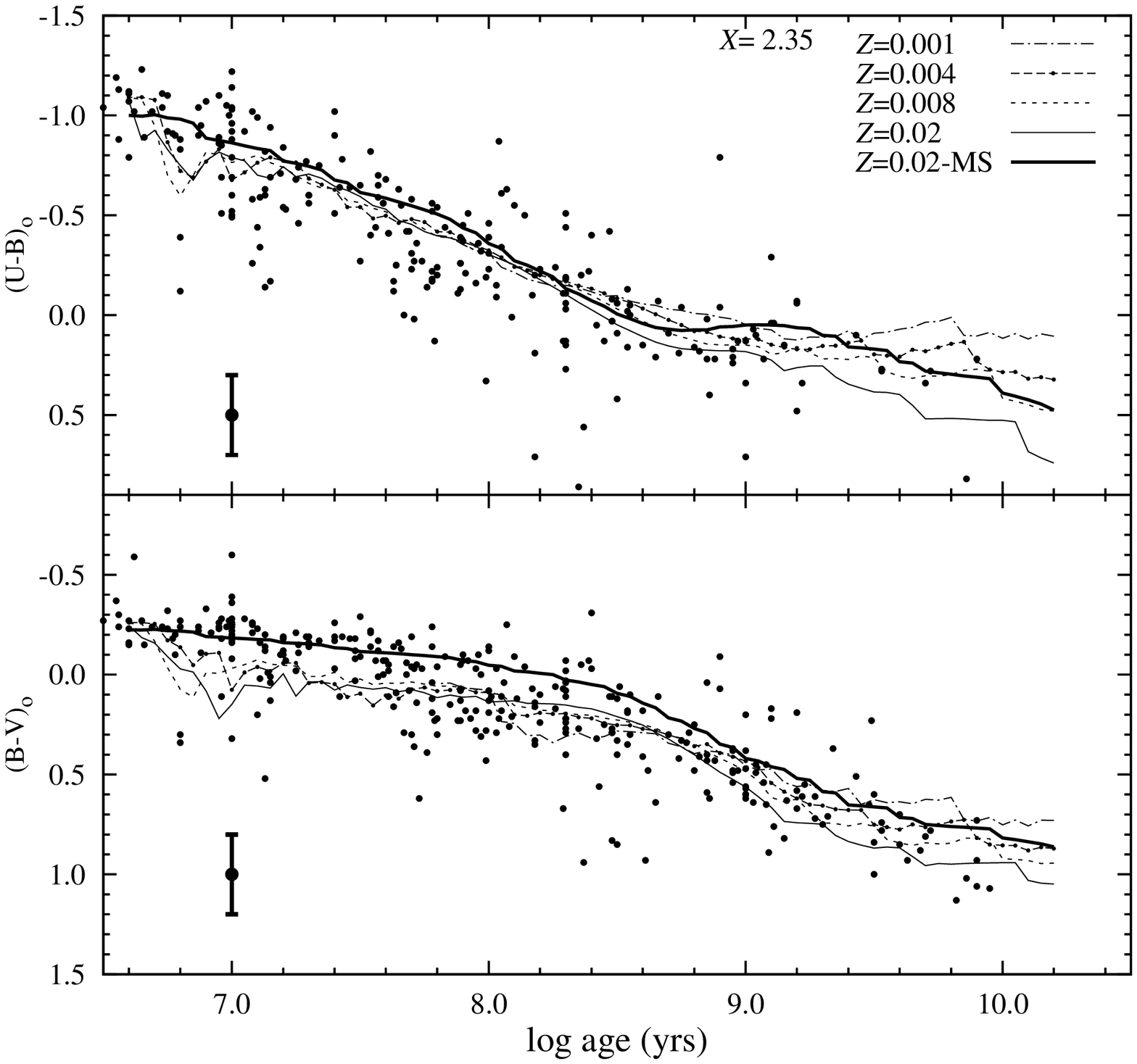}}

Figure 12(a): A comparison of present model predictions for whole
cluster population (dashed and thin curves) and MS population (thick
curve) with the observational data (whole population) for galactic
open clusters. The typical errors in estimation of integrated
colours are also shown. 
\end{figure*}

\textbf{}%
\begin{figure*}
\textbf{\includegraphics[angle=270,width=12cm]{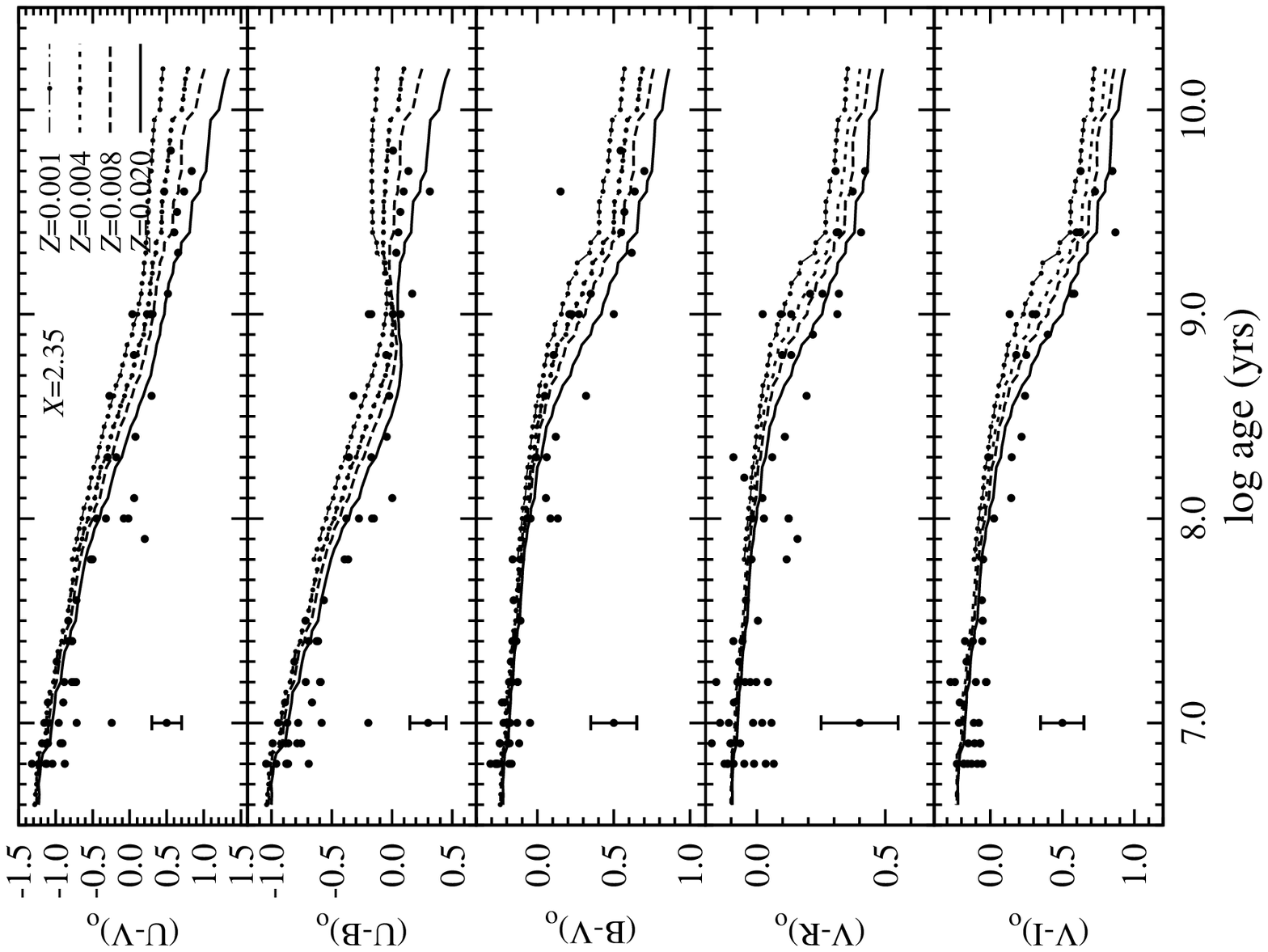} }

\caption{(b) Same \label{fig:12}as Figure 12a but for MS population only.
The typical errors in estimation of integrated colours are also shown.}

\end{figure*}

In Fig. 12(a) we compare evolution of $(U-B)$ and $(B-V)$
colours of a synthetic cluster ($X$ = 2.35, $Z$ = 0.02) having MS
as well as evolved population with the observed integrated parameters
of open clusters. Observed integrated parameters for 319 clusters
(whole population) have been taken from Lata et al. (2002). Although
the observational data show a large scatter, the agreement between
the theoretical and observed evolution of colours is good. Observed
integrated $(B-V)$ colours of clusters having log (age) $\leq$ 7.5
are better explained by MS model. Figure 12(b) shows a
comparison of integrated colours of MS population only, which indicates
a good agreement between the theoretical and observed colours.

\subsection{LMC clusters}

The comparison of $(U-B)$ and $(B-V)$ colour evolution by Bica et
al. (1996) with the present model is given in %
\begin{figure*}
\includegraphics[width=15cm]{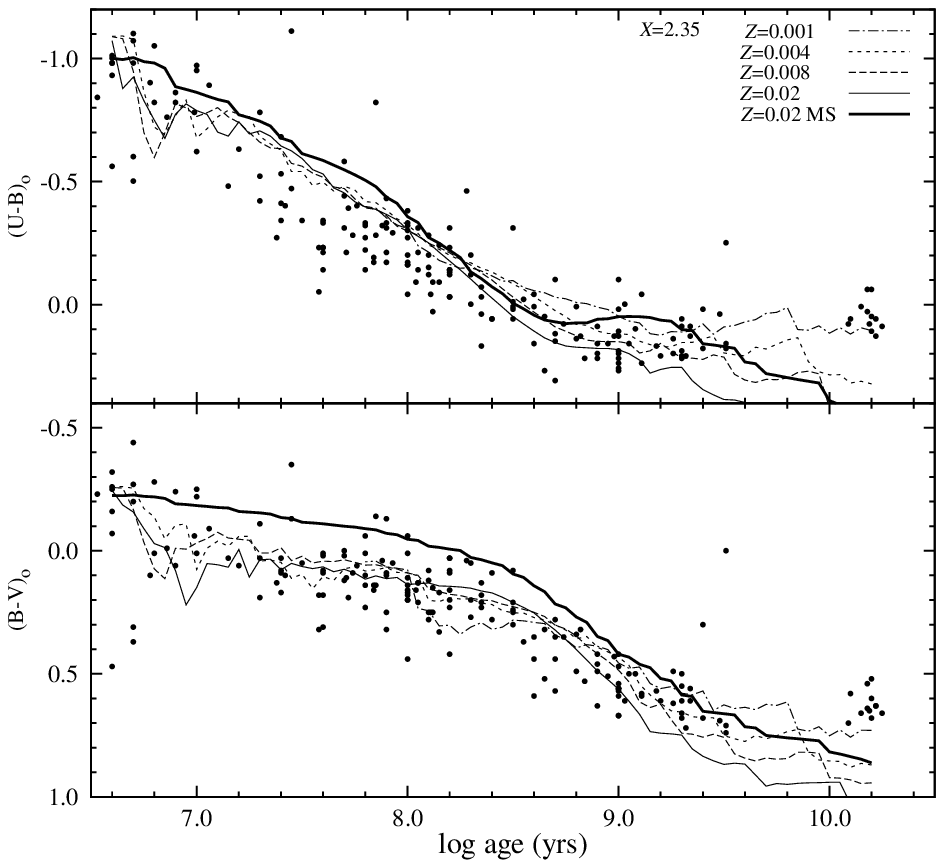} 

\textbf{\caption{Comparison \label{fig:13} of the present model (dashed and thin curves:
whole population; thick curve: MS population) with the observational
data for LMC cluster (whole population) by Bica et al. (1996).}
}
\end{figure*}
Fig. 13. The age of the clusters is taken from Sagar \&
Pandey (1989) and Mackey \& Gilmore (2003). A constant reddening of
$E(B-V)$ = 0.10 mag is applied to the observed data. The comparison
between observed data and the model indicates a fair agreement. A
few clusters having log (age) $\leq$ 8.0 are relatively bluer and
can be explained by the MS model. Figure 13 further confirms
the well known fact (cf. Olszewski et al. 1991, Olszewski et al. 1996) that the oldest clusters (age $>$ 10 Gyr) in the LMC are significantly
metal poor.%
\begin{figure*}
\includegraphics[width=14cm,angle=270]{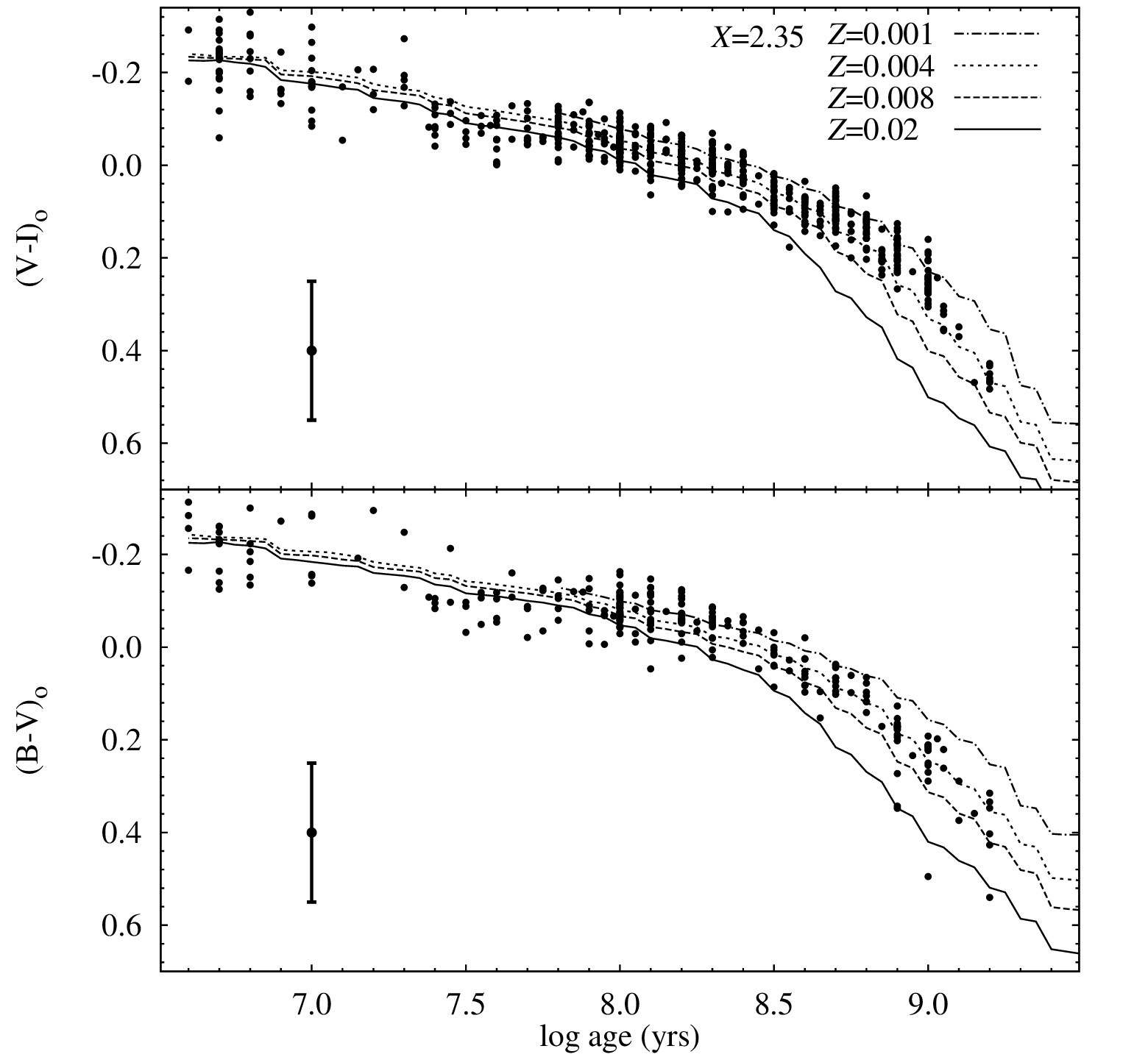} 

\textbf{\caption{Comparison\label{fig:14} of observed integrated colours for MS population
of LMC clusters obtained in the present work with the present model
predictions. The typical error in estimation of integrated
colours is also shown.}
}
\end{figure*}

In Fig. 14, integrated $(B-V)$ and $(V-I)$ colour evolution
of the MS population obtained in the present work is compared with
the colour evolution obtained for synthetic clusters. As can be seen
the MS population colours are not affected by the stochastic fluctuations.
The comparison indicates a nice agreement between the observed and
synthetic model colour evolution manifesting that the clusters having
age $>$ 500 Myr are distributed around lower metallicity ($Z$ =
0.004) model. However, the observed $(V-I)$ colours for clusters
having log (age) $>$ 8.7 are found to be bluer even than those for
models with $Z$ = 0.004.

\subsection{SMC clusters}

RZ05 compared their observational data with the models by Leitherer
et al. (1999) and Anders \& Fritze-v. Alvensleben (2003) and found
a systematic difference between their data and the models in the sense
that the observed data are too blue for the bluest colours. A large
scatter can be seen in the case of two colour diagrams namely $(U-B)$
vs $(B-V)$, $(V-I$) vs $(B-V)$ and $(U-B)$ vs $(V-I)$ diagrams.

\begin{figure*}
\includegraphics[angle=270,width=12cm]{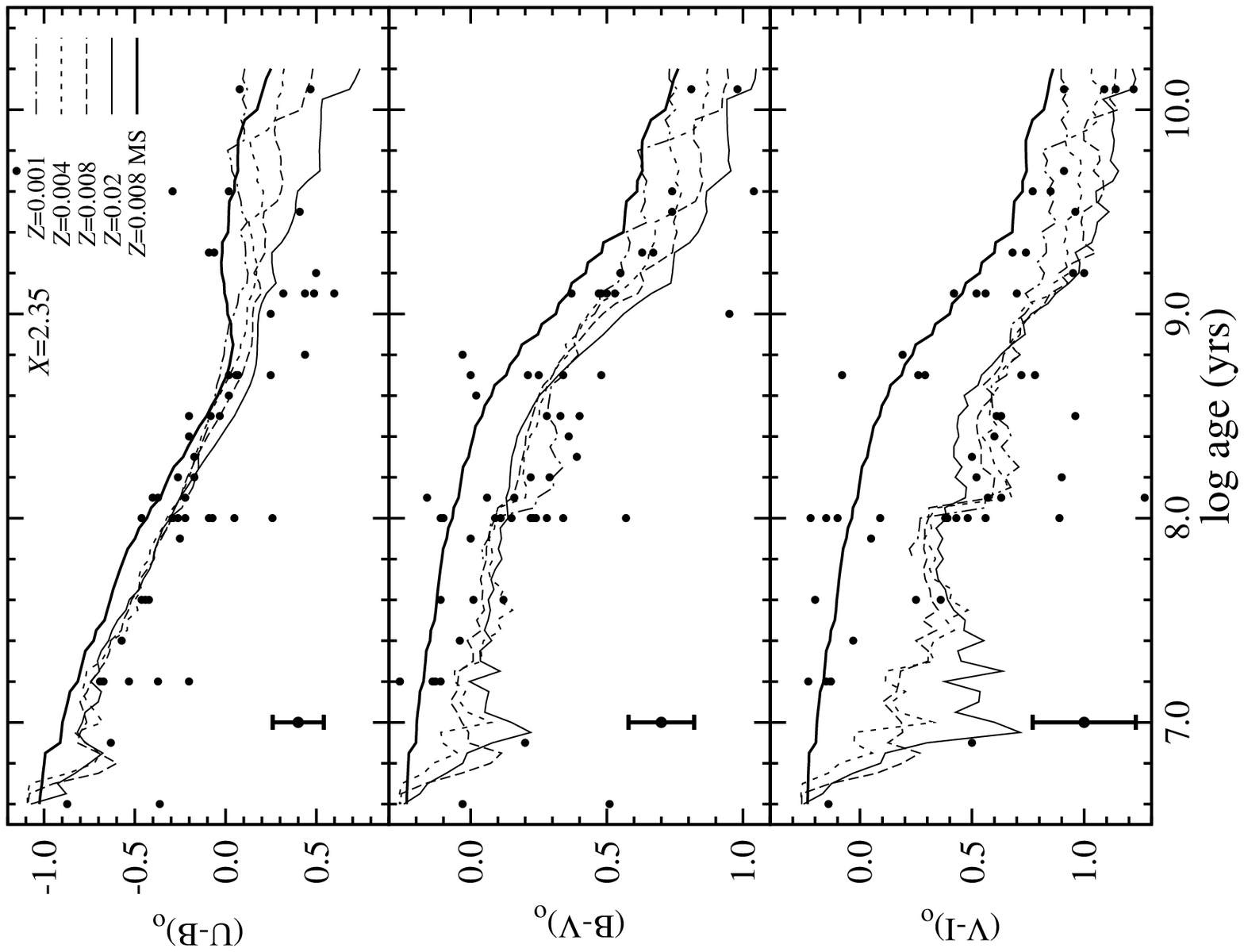} 

\textbf{\caption{Comparison of observational data by \label{fig:15}RZ05 with the present
model. The dashed and thin curves represent the whole population and
thick curve represents the MS population of the clusters. The
typical errors in estimation of integrated colours are also shown.}
}
\end{figure*}

In Fig. 15 we compare $(U-B)$, $(B-V),$ $(V-I)$ colour
evolution of SMC clusters using the data of RZ05 with the present
model. For comparison we assumed a mean reddening $E(B-V)$ = 0.1
mag. Comparison manifests that for the age range 6.5 $\leq$ log (age)
$\leq$ 8.0, the integrated $(B-V)$ and $(V-I)$ colours follow the
MS population colour evolution. The same trend has also been noticed
in the case of galactic open clusters and LMC clusters. Some of the
clusters in the age range log (age) $>$ 7.5 follow the whole cluster
population sequence predicted by the present model. The observed $(U-B)$
colour evolution is fairly represented by the present model colour
evolution.

\begin{figure*}
\includegraphics[width=14cm,angle=270]{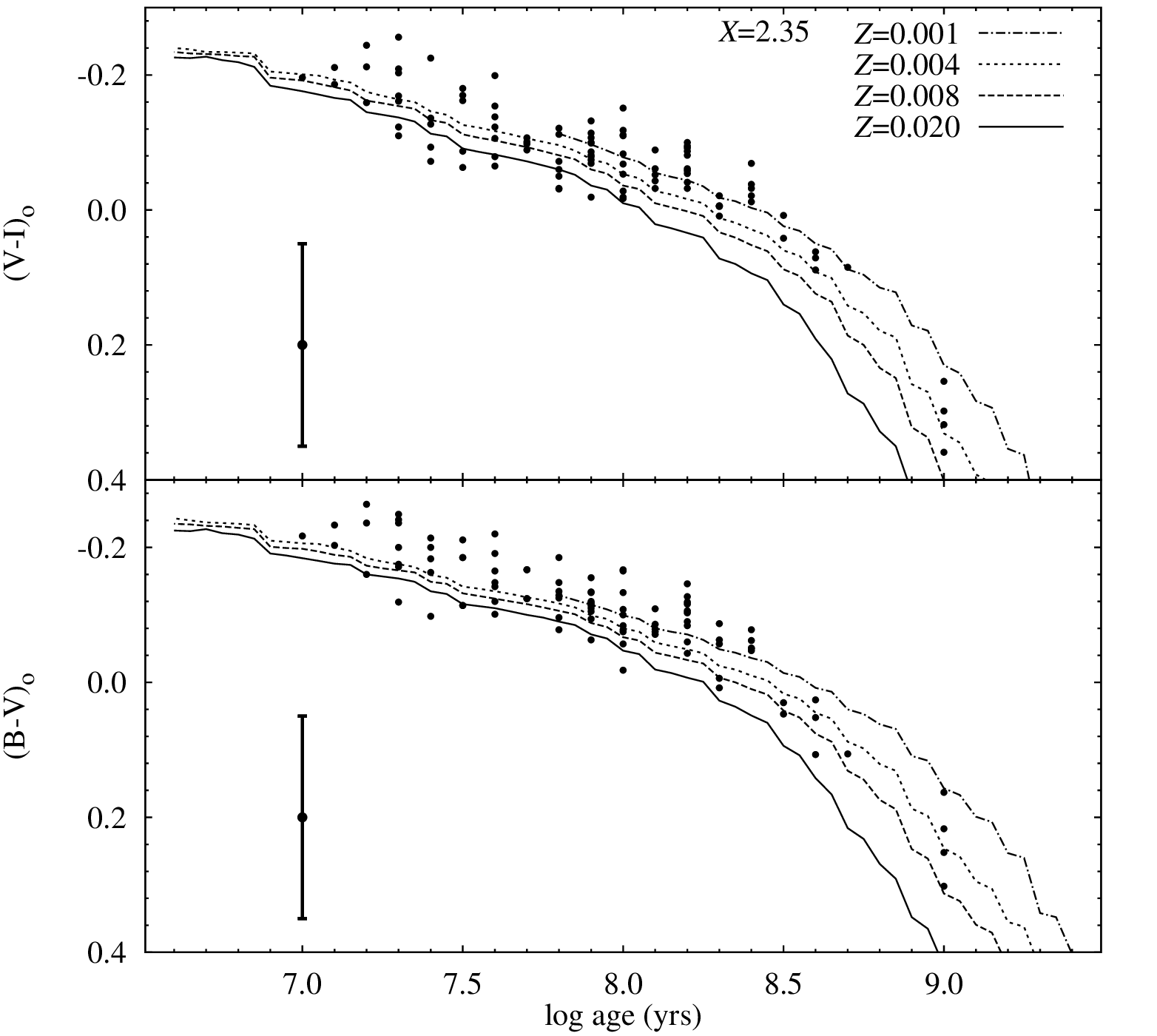} 

\textbf{\caption{Comparison of observed integrated colours for MS population of SMC
clusters \label{fig:16}with the present model predictions.
The typical error in estimation of integrated colours is also shown.}
}
\end{figure*}

Figure shows $(B-V)$ and $(V-I)$ colour evolution
of MS population of SMC star clusters. The model for $Z$ = 0.001,
0.004, 0.008 and 0.02 are also shown in the figure. The scatter in
the observational data is less just like in the case of open and LMC
clusters. The observed colour evolution in the age range log (age)
$\le$ 8.0 is fairly explained by the model having $Z$ = 0.004 and
0.008, whereas clusters having log (age) $>$ 8.0 seem to follow a
lower metallicity ($Z$ = 0.004) model.

\subsection{\textmd{Effect of stochastic fluctions on observed MC clusters}}

Girardi \& Bica (1993) have pointed out that the small number of evolved
red stars in less populous clusters can cause significant and fast
change in their integrated colours. They pointed out that most of
the dispersion in the observed $(U-B)$ {\it vs} $(B-V)$ diagram of LMC star
clusters (their figure 3) can be attributed to the stochastic effects,
specially for clusters older than $\sim$ 50 Myr, for which the internal
reddening is expected to be negligible. Girardi et al. (1995) concluded
that in a sample with such a low luminosity clusters as that of Bica
et al. (1996), the stochastic effects play a significantly role in
interpretation of evolution of integrated clusters.

\textbf{}%
\begin{figure}
\textbf{\includegraphics[angle=270,origin=c,width=12cm]{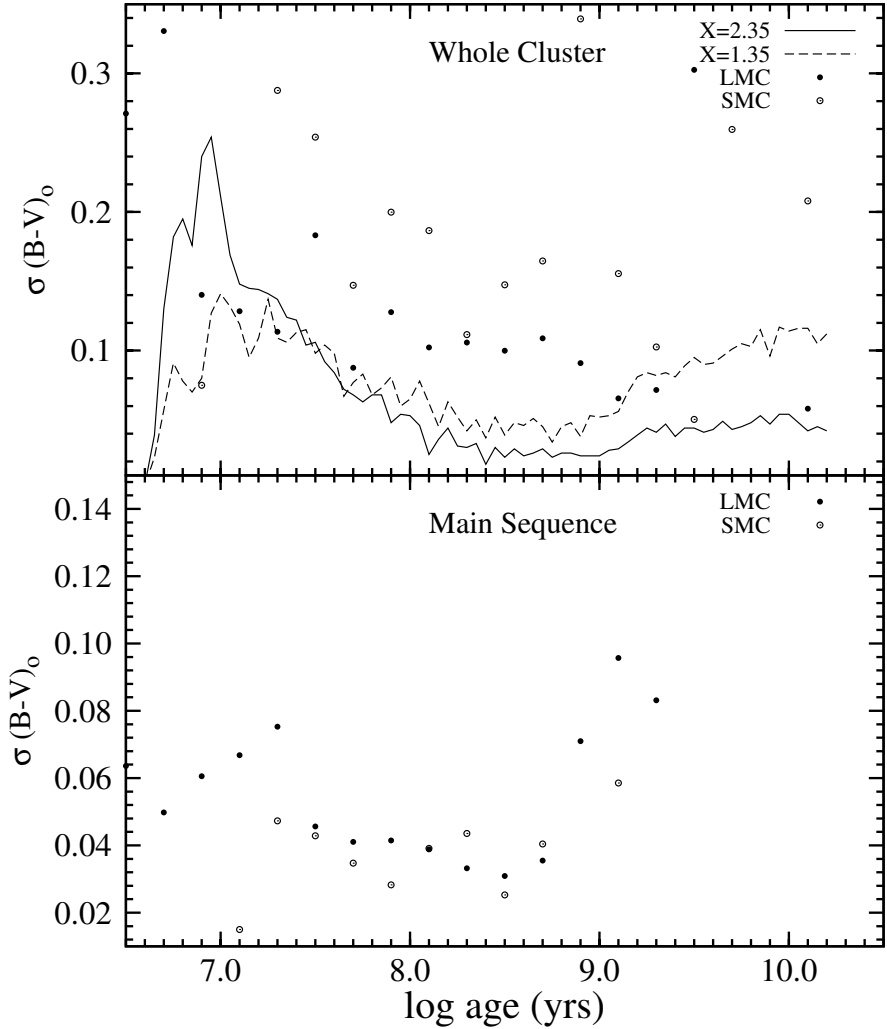}}

\caption{Effect of stochastic fluctuations on observed colours of MC star clusters (filled circles: LMC clusters; open circles: SMC clusters). The continuous and dashed curves show effect of stochastic fluctuations on colours of simulated clusters having post MS population.}

\end{figure}

As Girardi \& Bica (1993) pointed out that the dispersion in the observed
colour-colour and colour evolution diagram is mainly due to stochastic
effects, we used Figs 13 - 16 to study the effect of stochastic fluctuations
on observed $(B-V)$ colours of MC star clusters. Figure 17 shows
standard deviations of mean $(B-V)$ colours in a bin of $log (age)$
= 0.2 as a function of age, which clearly shows that the dispersion
in the case of MS population is significantly less than in the case
of whole cluster population.

\subsection{\textmd{Age-metallicity relation for MC star clusters}}

The age-metallicity relation in the MC star clusters is well known
since a long time. For example, a compiled catalogue by Sagar \& Pandey
(1989) yields 0.001 $\leq Z\leq$ 0.01 with a mean value of $Z\sim0.005$
for cluster having log age $\sim$ 7.0 - 9.2 (cf. their figure 3),
whereas the study of LMC clusters by Olszewski et al. (1991) indicates
a mean value of $Z\sim0.008$ with a range of $Z\sim$ 0.02 - 0.004.
Bica et al. (1998) have derived mean metallicity for the intermediate
age LMC clusters ($9.0\leq log (age) \leq9.4$) as $z\sim0.005$ and
found that the metallicities obtained by them are significantly lower
than those reported by Olszewski et al. (1991) for a sample of clusters
of similar age, but their values are in good agreement with several
contemporary studies. In a recent study Kerber et al. ( 2007) have
found that the LMC clusters younger than $log (age) \sim9.5$ have $z\sim0.006$
with a considerable scatter. They have also pointed out that the metallicities
by Olszewski et al. (1991) are higher as compared to their values
and have also discussed possible reasons for such a  discrepancy. Above
discussions indicate that the colour evolution of MC star clusters
discussed in Secs. 4.2 and 4.3 is in accordance with the observed
age-metallicity relation for MC star clusters. However a discrepancy
in $(V-I)$ colours for LMC clusters having $log (age)\geq8.7$ has been
noticed in Sec. 4.2.

\subsection{LMC clusters: $(V-I)$ colour discrepancy}

Possible reasons for the discrepancy in $(V-I)$ colours of LMC clusters as 
noticed in Sec. 4.2 may be: i) reddening correction, ii) systematic effects 
in the model predictions towards older ages, and iii) anomalous reddening law.

The reddening $E(B-V)$ can be estimated relatively accurately for
Galactic open clusters, hence reddening corrections for individual
clusters were applied. Figure 12b shows a fair agreement of observed
colour evolution of MS population of Galactic open clusters with the
model predictions having $Z=0.02$, which suggests that the model
predictions do not have any systematic effect. In the case of MC star
clusters, we applied mean values of $E(B-V)$ for three age groups
(cf. Sec. 3.2) assuming a normal reddening law. In the case of LMC
clusters having $log (age) \geq8.7$ a mean value of $E(B-V=0.03)$ is
applied. As discussed above the $(B-V)$ colour evolution of LMC clusters
fairly agrees with the model predictions. Above facts indicate that
the systematic effects in model predictions and reddening correction
should not be the possible reasons for the discrepancy in $(V-I)$
colours.

\section{Two colour diagrams}

\label{sec:Two-colour-diagrams 5}

\subsection{\textmd{$(V-I)$ vs $(B-V)$ diagram}}

\begin{figure*}
\includegraphics[width=16cm,angle=270]{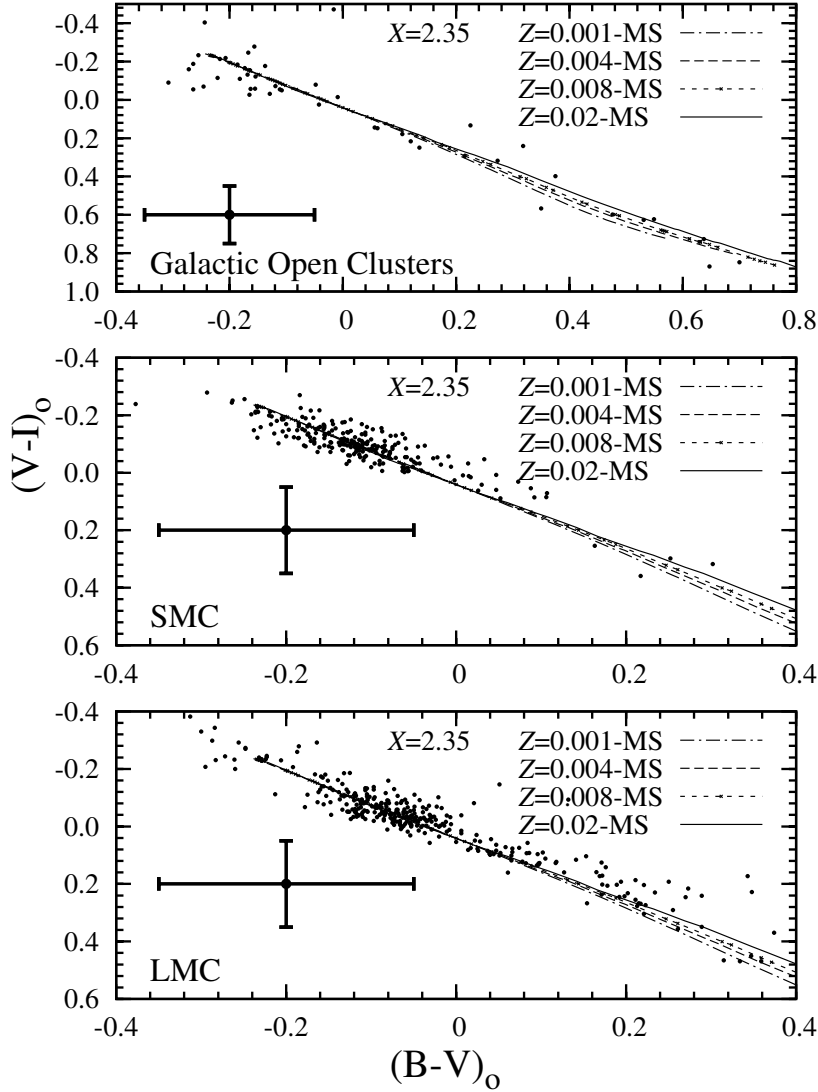} 

\textbf{\caption{(a) $(B-V)_{0}$ vs $(V-I)_{0}$ two colour diagram for MS population
of galactic open clusters (upper panel), SMC clusters (middle panel)
and LMC clusters (lower panel) compared with the present model.The
typical errors in estimation of integrated colours are also shown.}
}
\end{figure*}

\begin{figure*}
\includegraphics[width=18cm,angle=270]{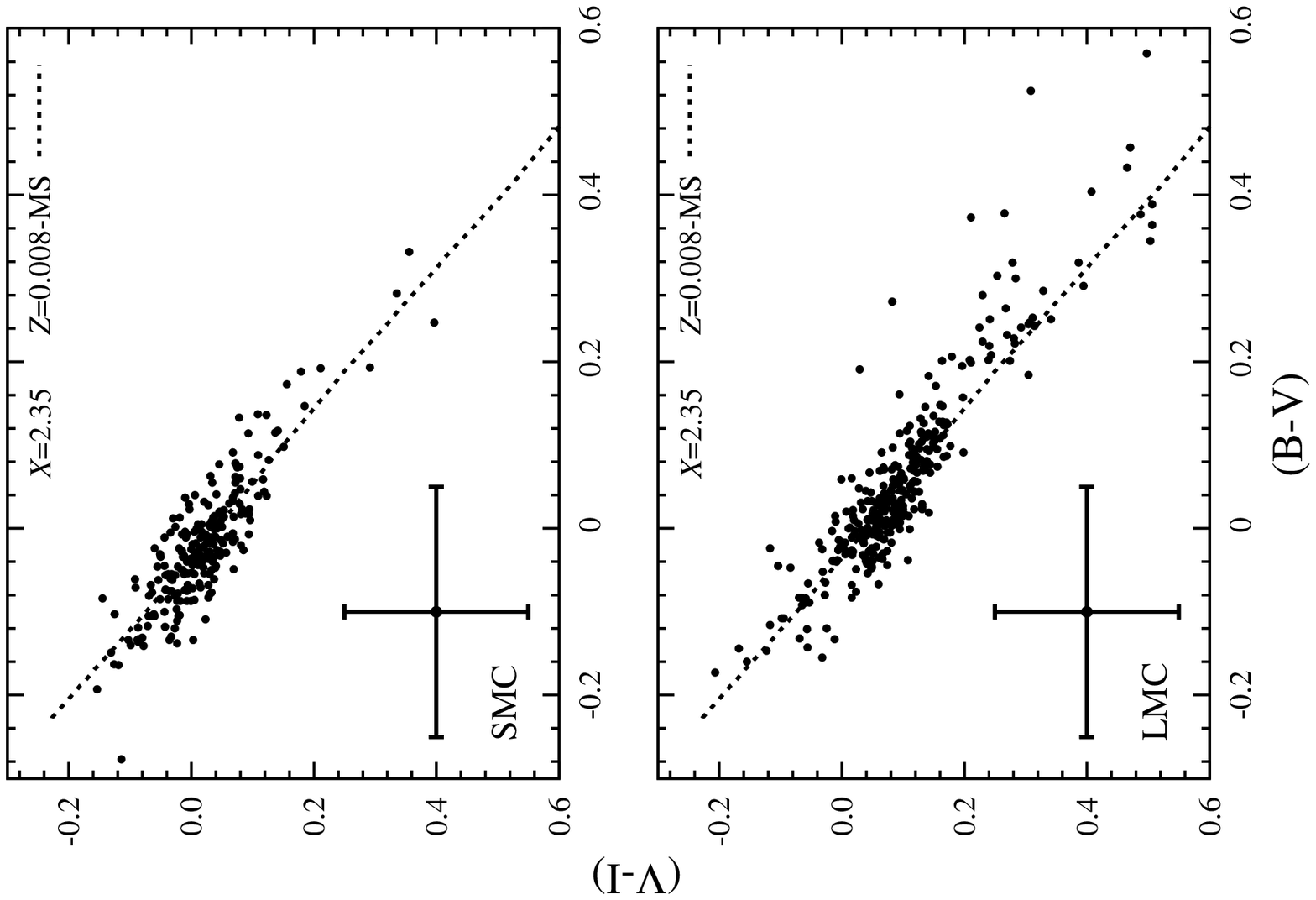} 

Figure 18(b): $(B-V)$ vs $(V-I)$ two colour diagram for MS population
SMC clusters (upper panel) and LMC clusters (lower panel) compared
with the present model with $Z=0.008$. The typical errors
in estimation of integrated colours are also shown.
\end{figure*}

\begin{figure}

\includegraphics[angle=270,width=15cm]{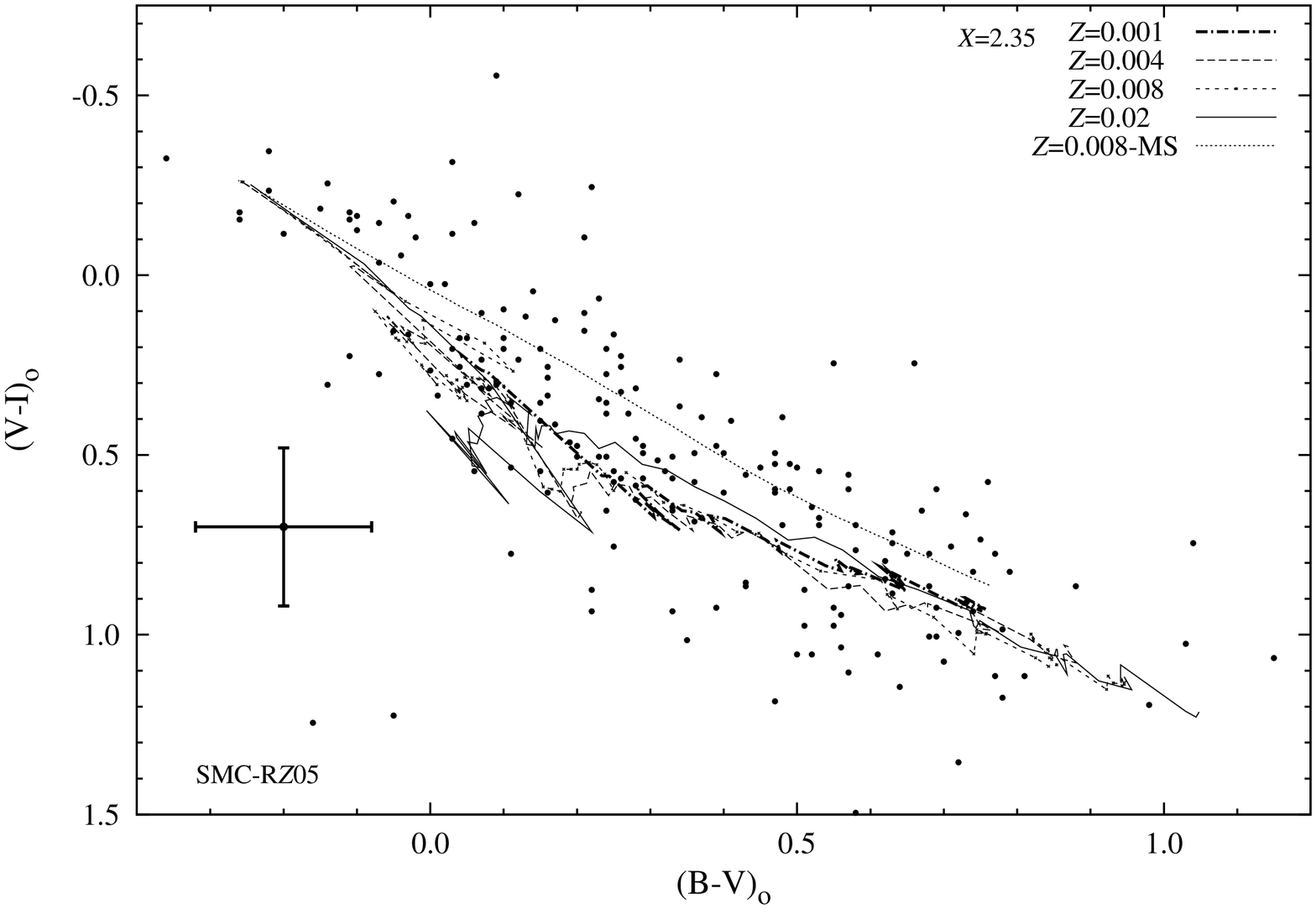}

\caption{Integrated colours of SMC clusters by RZ05 are compared with the present
model (dashed and thin curves: whole population; thick curve; MS population).
The typical errors in estimation of integrated colours are
also shown.}

\end{figure}

Figure 18a shows integrated $(V-I)_{0}$ vs $(B-V)_{0}$ two-colour
diagram (TCD), assuming a normal reddening law, for galactic open
clusters, LMC clusters and SMC clusters along with the present model
predictions. Figure 18a indicates that the model nicely explains observed
colours of galactic open clusters. The MS population colours of LMC
and SMC star clusters having $(B-V)>$ 0.0 become sytematically red
in $(B-V)$ colour or blue in $(V-I)$ colour. Figure 18b shows reddened
$(V-I)$ vs $(B-V)$ TCDs along with model predictions for $Z=0.008$
for MC star clusters, which also shows discrepancy in the colours.

This anomaly may be due to anomalous reddening law in the MC. The
$(B-V)$ vs $(V-I)$ diagram can be used to study the reddening law
(cf. Pandey et al. 2003). A least-square fit to the observed LMC and
SMC data shown in Fig. 18b gives a slope $m_{MC}=0.9\pm0.04$, whereas
a fit to the model ($Z=0.02$ and 0.008) yields a slope $m_{normal}=1.08\pm0.01$
and $m_{normal}=1.12\pm0.01$ respectively. Adopting the
procedure described by Pandey et al. (2003), the value of total to
selective absorption $R_{V(MC)}$ towards MC can be obtained as follows:

\[
R_{V(MC)}\simeq\frac{m_{MC}}{m_{normal}}\,\times\, R_{V}\]
 assuming $R_{V}$= 3.1, the value of $R_{V(MC)}$ comes out to be
2.6 - 2.5 $\pm0.1$ indicating a lower grain size towards the MC star
clusters.

There are evidence for the anomalous reddening law in the Magellanic
Clouds. For the SMC bar Gordon \& Clayton (1998) and Gordon et al.
(2003) have found $R_{V}=2.7\pm0.1$, which is consistent with the
value $2.7\pm0.2$ reported by Bouchet et al. (1985). Smaller value
of $R_{V}=2.76\pm0.09$ was reported for the LMC2 super-shell sample
by Gordon et al. (2003). However, for the LMC average sample Gordon
et al. (2003) have found $R_{V}=3.41\pm0.06$.

The metallicity in the Magellanic Clouds is substantially lower than
in the Milky Way and there are indication that measured extinction
curves towards LMC and SMC differ from typical extinction curves in
the Milky Way (cf. Weingartner \& and Draine 2001). Because of lower
metallicity the typical molecular clouds in the LMC and SMC are bigger
but more diffuse than those in the Milky Way (Pak et al. 1998). Therefore,
dust grains in the LMC and SMC may not spend as much time in dark,
shielded environment as dust grains in the case of Milky Way. This
may result to small size dust grains, which consequently yield low
value of $R_{V}$ in the in the LMC and SMC.

A comparison of RZ05 data with the present model (Fig. 19) indicates
that majority of the observed data is fairly explained by the MS population,
however some of the observations follow the whole cluster model. A
comparison of Figs 18a and Fig. 19 also indicates that the scatter
in MS population data is significantly less than the data of RZ05.

\subsection{\textmd{$(U-B)$ }vs $(B-V)$ diagram}

\begin{figure*}
\includegraphics[angle=270,width=12cm]{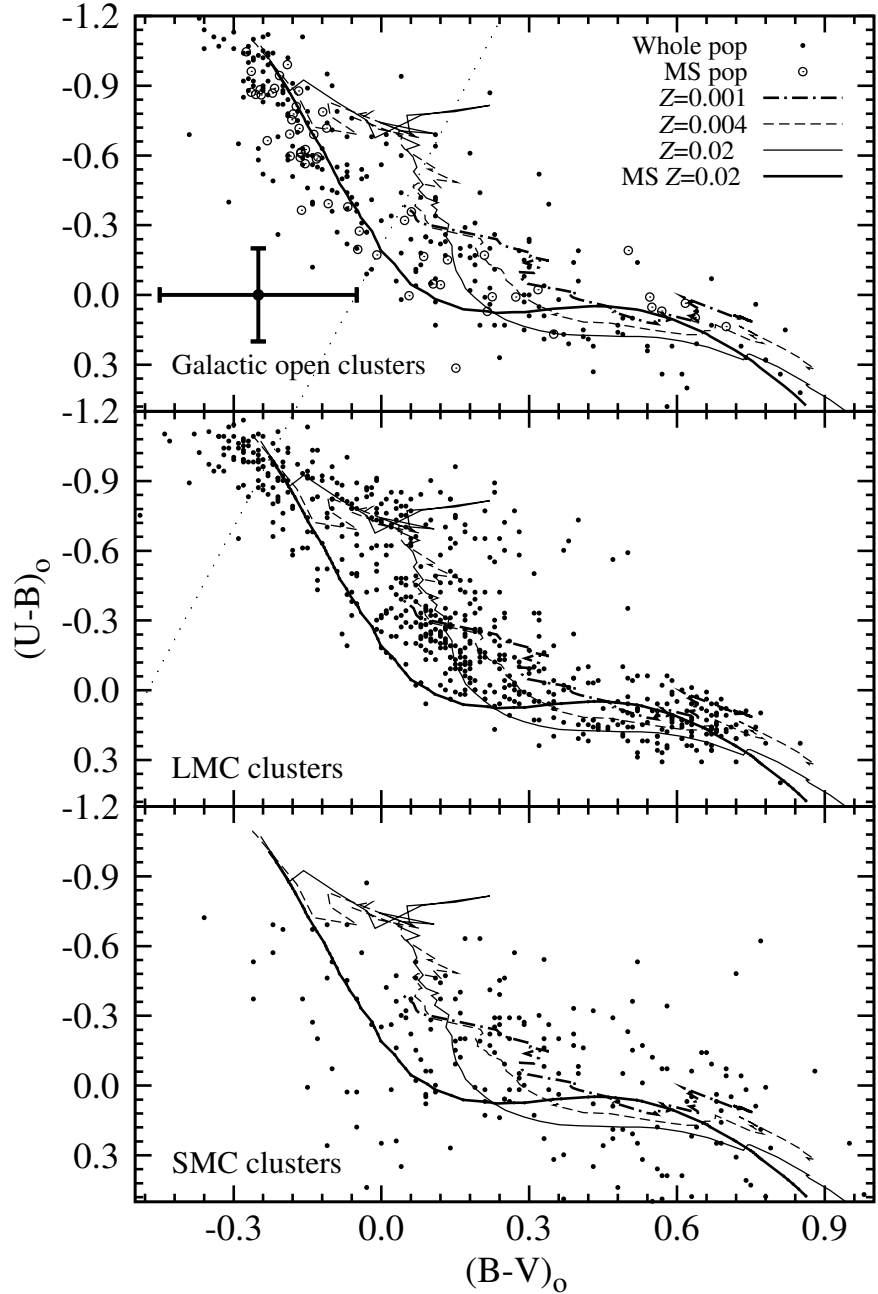}

\textbf{\caption{$(U-B)_{0}$ vs $(B-V)_{0}$ synthetic colour-colour diagram compared
with the observational data of Galactic open clusters (whole population;
Pandey at al. 1989 and Lata et al. 2002, MS population: present work),
LMC (Bica et al. 1996) and SMC (RZ05) clusters. }
}
\end{figure*}

Figure 20 shows $(U-B)_{0}$ vs $(B-V)_{0}$ two colour diagram for
MS population ($Z=$0.02, $X=$-2.35) and whole clusters population
of a synthetic cluster ($Z$ = 0.004, 0.008, 0.02, $X=$-2.35) and
compares it with the observational data of galactic open clusters,
LMC and SMC clusters. In the case of LMC data (Bica et al. 1992, 1996)
and SMC data (RZ05) a mean reddening of $E(B-V)$ = 0.1 mag has been
assumed.

Figure 20 shows a large scatter in the observational data. The amount
of scatter is almost the same in various sets of data. In all the
three samples a large number of the bluest clusters ($(B-V)_{o}<$
0.0) follow the MS population relation. Note that similar trend in
the age range 6.5 $\leq$ log (age) $\leq$ 8.0 has been observed
while studying the $(B-V)$ colour evolution (cf. Sect 4).
Remaining clusters show a large scatter around the theoretical $(U-B)$
vs $(B-V)$ relation for the whole cluster.

\section{\textmd{Conclusion}}

\label{sec:Conclusion 6}

In this paper we present integrated magnitude and colours for synthetic
clusters using the synthetic CMDs of star clusters. The integrated
parameters have been obtained for the whole cluster population as
well as for the MS population of star clusters. We have also estimated
observed integrated magnitude and colours of MS population of galactic
open clusters, LMC and SMC star clusters. The relation between observed
integrated colours for whole cluster population and MS population
are fairly explained by the model predictions obtained in the present
work. This indicates that the estimated observed integrated colours
of MS population of MC clusters fairly represent the MS population
of the MC clusters. Main conclusion of the present study are;

\begin{enumerate}
\item Present model suggests that colour evolution of MS population of star
clusters is not affected by the stochastic fluctuations. Stochastic
fluctuations significantly affect the colour evolution of the whole
cluster population. The fluctuations are maximum in $(V-I)$ colour
in the age range 6.7 $<$ log (age) $<$ 7.5. The observed data of
MC star clusters also indicate that the effect of stochastic fluctuations
on estimation of integrated colours of MS population is significantly
less than in the case of colours of whole population.
\item The evolution of integrated magnitude of star clusters with the age
depends on the IMF of the cluster. Presence of massive stars, i.e.
shallow IMF, makes the integrated magnitude of cluster brighter that
fades relatively faster than the clusters having steeper IMF. Variation
of IMF has insignificant effect on colour evolution of star clusters
after log (age) $\sim$ 7.5. However the colour evolution in the age
range 6.7 $<$ log (age) $<$ 7.5 is significantly governed by the
choice of the IMF. This further confirms the earlier results e.g.
by Chiosi et al. (1988), Pandey et al. (1989), Girardi et al. (1995),
Bruzual \& Charlot (2003) and references therein. 
\item The metallicity variation does not show any significant effect on
the evolution of magnitude as well as on the colours of clusters having
log (age) $\leq$ 8.5. For older clusters colours become bluer with
the decrease in metallicity.  This is in accordance with the results
obtained in earlier studies (e.g. Girardi et al. 1995). 
\item The $(U-B)$ and $(U-V)$ colours for whole cluster population are
slightly bluer in comparison to those reported by Brocato et al. (1999)
and Maraston et al. (1998). 
\item The $(B-V)$ colour evolution for the whole cluster population is
in agreement with those reported by Brocato et al. (1999) and Maraston
et al. (1998). The $(V-I)$ colour evolution for the log (age) $\geq$
7.0 is in reasonable agreement with that given by Brocato et al. (1999). 
\item The $(V-R)$ colour evolution obtained in the present work is in good
agreement with that given by Brocato et al. (1999), whereas the $(V-R)$
colour evolution reported by Maraston et al. (1998) does not agree
with the present work as well as with that given by Brocato et al.
(1999). 
\item Evolution of integrated colours of MS population of the clusters in
the Milky Way, LMC and SMC obtained in the present study are nicely
explained by the present synthetic cluster model. A comparison of
present model with the observational data indicates that the MC star
clusters having age $\geq$ 500 Myr  seems to favour a metallicity lower than 
$Z=$0.008. Observed integrated colours of the whole population of
Milky Way, LMC and SMC star clusters are also explained fairly well
by the present model. 
\item $(V-I)$ vs $(B-V)$ two-colour diagram for the MS population of the
Milky Way star clusters shows a fair agreement between the observations
and present model. However, the diagrams for LMC and SMC indicate a
discrepancy in colours. An anomalous reddening law towards the MC
may be a possible reason for the discrepancy.
\item Comparison of synthetic $(U-B)$ vs $(B-V)$ relation with the observed
data (i.e., whole cluster population) of Milky Way, LMC and SMC star
clusters indicates that majority of the bluest clusters ($(B-V)_{o}<$
0.0) follow the MS population relation. The observed colour evolution
of young cluster (6.5 $\leq$ log (age) $\leq$ 8.0) in the Milky
Way, LMC and SMC also indicates that a large number of young clusters
follow the MS population relation. 
\item The $(U-B)$ vs $(B-V)$ colour-colour diagram and colour evolution
of star clusters are frequently being used to date the clusters by
comparing observed data with the model prediction for whole cluster
population. Present results indicate that the dating of the clusters
may be erroneous if proper synthetic model  (e.g. whole population
model is used with out proper statistical techniques to account for
the stochastic fluctuations (see Bruzual 2009)) is not used.
\end{enumerate}

\section{\textmd{ACNKOWLEDGMENTS}}

We are thankful to the anonymous referee for the critical comments
which improved the scientific contents and presentation of the paper.

\end{document}